\documentclass[pra,twocolumn,citeautoscript,superscriptaddress]{revtex4-1}

\usepackage{hyperref}
\usepackage{grffile}
\usepackage{graphicx}
\usepackage{amsmath}
\usepackage{amssymb}
\usepackage{color}
\usepackage{mathrsfs}
\usepackage{braket}

\begin{document}

\title{Spin transport in a one-dimensional quantum wire}

\author{A.-M. Visuri}
\email{anne-maria.visuri@unige.ch}
\affiliation{Department of Quantum Matter Physics, University of Geneva, 24 quai Ernest-Ansermet, 1211 Geneva, Switzerland}
\author{M. Lebrat}
\affiliation{Department  of  Physics,  ETH  Zurich,  8093  Z\"urich,  Switzerland}
\affiliation{Department of Physics, Harvard University, 17 Oxford St, Cambridge, MA, 02138, USA}
\author{S. H\"ausler}
\affiliation{Department  of  Physics,  ETH  Zurich,  8093  Z\"urich,  Switzerland}
\author{L. Corman}
\affiliation{Department  of  Physics,  ETH  Zurich,  8093  Z\"urich,  Switzerland}
\author{T. Giamarchi}
\affiliation{Department of Quantum Matter Physics, University of Geneva, 24 quai Ernest-Ansermet, 1211 Geneva, Switzerland}

\begin{abstract}

We analyze the spin transport through a finite-size one-dimensional interacting wire connected to noninteracting leads. By combining renormalization-group arguments with other analytic considerations such as the memory function technique and instanton tunneling, we find the temperature dependence of the spin conductance in different parameter regimes in terms of interactions and the wire length. The temperature dependence is found to be nonmonotonic in part of the parameter space. In particular, the system approaches perfect spin conductance at zero temperature for both attractive and repulsive interactions, in contrast with the static spin conductivity. 
We discuss the connection of our results to recent experiments with ultracold atoms and compare the theoretical prediction to experimental data in the parameter regime where temperature is the largest energy scale.

\end{abstract}

\maketitle

\section{Introduction}

The transport of spin, instead of electric charge, is the basis of the field of spintronics~\cite{Zutic_spintronics2004}. Devices utilizing spin current can potentially be made more compact and consume less power than electronics devices. Some such devices are already in use, including magnetic memories and magnetic field sensors, and various new materials are currently investigated for spintronics applications~\cite{Gong_heterostructure_devices2019}. Spin transport has raised interest also more generally, and has been probed for instance in experiments with ultracold atoms. Trapping ultracold atomic gases in optical potentials allows for studing the transport of spin in highly controllable and tunable environments~\cite{Gensemer_collisionless_to_hydrodynamic2001, Sommer_universal2011, Krinner_spin_and_particle2016, Luciuk_quatum-limited2017, Valtolina_spin_dynamics2017, Fava_spin_superfluidity2018, Nichols_Mott_insulator2019}. The internal states of atoms can be used to emulate the spin degree of freedom. 

Spin current is related in linear response to the application of a magnetic field gradient. In the case of free particles, spin current is perfectly conducted, with an infinite static conductivity, whereas interactions between particles in different spin states lead to spin diffusion and to a finite spin conductivity. For an infinite system, conductivities can be computed using the Kubo formula \cite{Mahan_many-particle2000}. For a mesoscopic system attached to reservoirs, on the other hand, formalisms
such as Landauer-B\"uttiker \cite{Landauer_spatial_variation1957, Landauer_electrical_resistance1970, Buttiker_multiprobe_conductors1988} or its generalizations to interacting particles \cite{Meir_interacting1992} are usually employed. In this case, the relevant quantity describing transport through the system is conductance. These transport quantities are affected by the interactions between particles.
The effect of interactions on conductance has been studied in various settings, in particular for an interacting wire connected to leads. In the case of a one-dimensional wire, conductance is known to be independent of interactions in the wire when the interactions are of the Tomonaga-Luttinger liquid type~\cite{Safi_transport1995, Maslov_conductance1995, Ponomarenko_conductance1995}. Other similar situations include the charge and spin transport through backscattering barriers~\cite{Safi_proprietes1997, Safi_dirty_wire1999} and periodic potentials~\cite{Ponomarenko_treshold_features1997, Ponomarenko_Mott_insulator1998, Ponomarenko_spin_gap_insulators2000, Ponomarenko_spin_transport_Mott2014, Morath_inhomogeneous_wires2016}, and the case of Coulomb interactions which lead to the formation of a Wigner crystal~\cite{Matveev_Wigner-crystal2004, Matveev_low_electron_density2004, Meyer_Wigner_crystal2008}.
 
A quantity related to conductance, spin drag, arises when interactions between particles in different spin states lead to a friction between the different spin components. Spin drag is analogous to the Coulomb drag between electrons~\cite{Narozhny_coulomb_drag2016}, demonstrated in experiments with electrons in two two-dimensional layers separated by a tunnel barrier. Spin drag was proposed to occur when the layer degree of freedom is replaced by the spin~\cite{DAmico_spin_coulomb_drag2000}, and was observed experimentally in a two-dimensional electron gas~\cite{Weber_spin_Coulomb_drag2005}. 
Spin drag reduces the total spin current and is therefore relevant for spintronics applications. In real materials, spin drag effects are screened by relaxation mechanisms such as scattering from phonons or impurities. Such mechanisms are not present in cold-atom experiments and the damping of spin currents is solely due to interactions. Spin drag has indeed been probed recently also in experiments with cold atoms~\cite{Sommer_universal2011, Koller_quantum_enhancement2015, Krinner_spin_and_particle2016}. In particular, a setup with a quantum point contact between two atom cloud reservoirs was used to measure spin and particle conductances and the spin drag~\cite{Krinner_spin_and_particle2016}. 

Motivated by these recent transport experiments with ultracold atoms \cite{Krinner_spin_and_particle2016, Lebrat_band_and_correlated2018}, we consider a finite one-dimensional quantum wire connected to leads. Given the one-dimensional nature of the problem, spin and charge are decoupled, and the charge sector is described by a Tomonaga-Luttinger liquid (TLL)~\cite{Giamarchi_one_dimension2003}. The charge conductance of a TLL wire of finite length can be calculated exactly~\cite{Safi_transport1995, Maslov_conductance1995, Ponomarenko_conductance1995} and was shown, in the case of noninteracting leads, to be equal to the conductance quantum. 
We consider here the case of spin transport in such a wire. For the spin sector, the presence of backscattering between opposite spins directly affects transport even if the wire is perfectly invariant by translation. We compute the spin conductance and spin drag as a function of the interactions, the length of the wire, and temperature. We discuss possible consequences for cold-atom experiments and, in particular, compare our results to experimental data which is available in the regime where temperature is the highest energy scale~\cite{Krinner_spin_and_particle2016}.
We use renormalization group to analyze the spin Hamiltonian in different parameter regimes, defined in terms of the wire length, superfluid coherence length, and thermal length. The temperature dependence of the spin conductance in these different regimes is found by perturbative calculations. One of the central results of the paper, shown in Fig.~\ref{fig:sketch}, is a nonmonotonic dependence of the spin conductance on temperature. 

In the following sections, Sec.~\ref{sec:model} introduces the microscopic model and the effective low-energy field-theory description as well as the quantities used for characterizing spin transport. Sections~\ref{sec:high_temperature} to \ref{sec:low_temperature} discuss the spin conductance at high, intermediate, and low temperature, respectively, based on renormalization group analyses and perturbative calculations. Finally, Sec.~\ref{sec:discussion} summarizes the behavior of the spin conductance in the different parameter regimes discussed in Sec.~\ref{sec:high_temperature}--\ref{sec:low_temperature}. Conclusions are presented in Sec.~\ref{sec:conclusions} and technical details in the appendices.

\section{Model}
\label{sec:model}

\subsection{Interacting particles in a quantum wire}

\begin{figure}[h]
\centering
\includegraphics[width=0.7\linewidth]{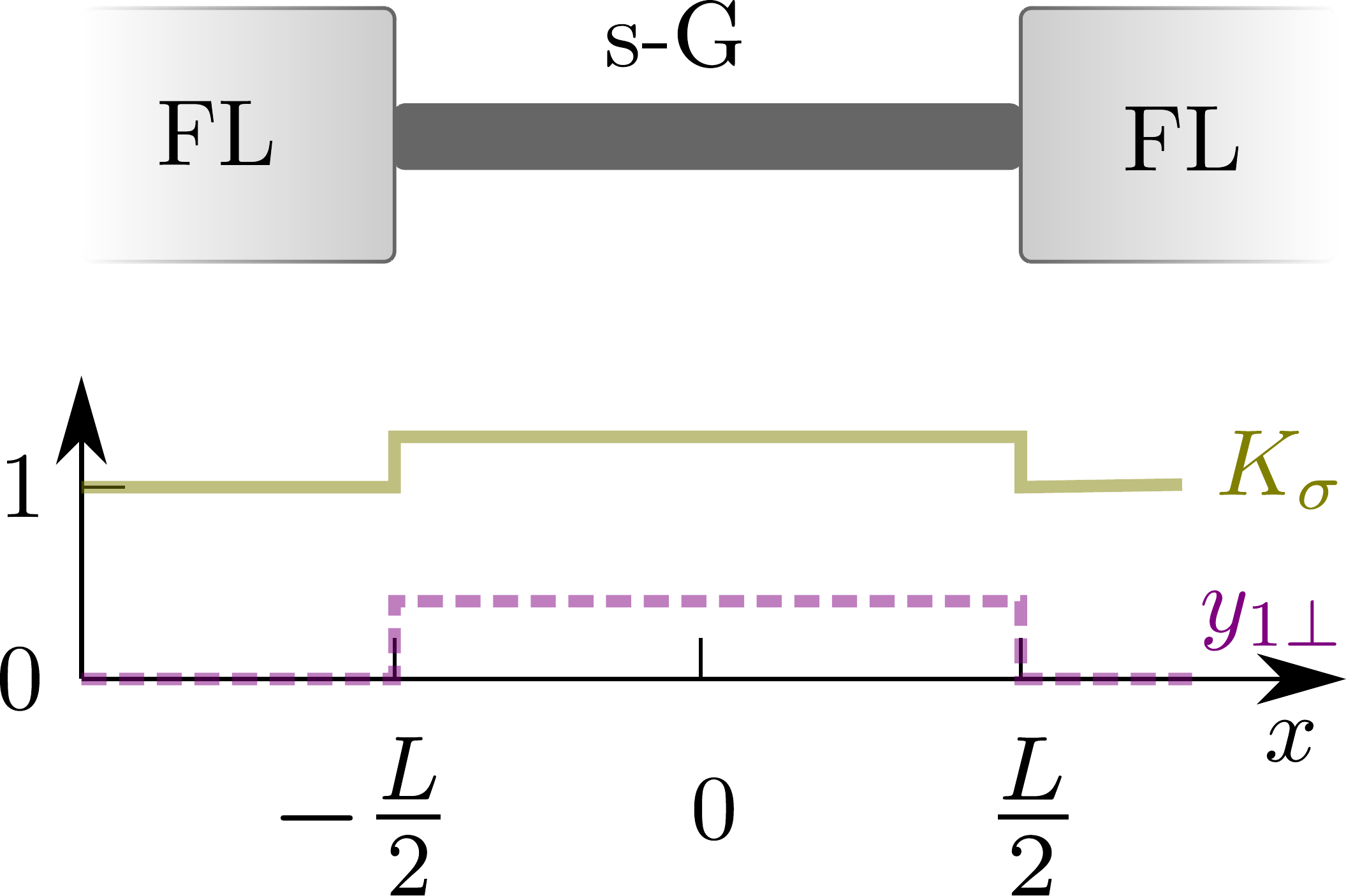}
\caption{The one-dimensional wire is coupled to Fermi-liquid (FL) leads, modeled by noninteracting 1D systems. The interacting wire is described by the sine-Gordon (s-G) model, as discussed in Sec.~\ref{sec:bosonization}. The wire region has a finite length $L$ whereas the leads are infinite. In the leads, $K_{\sigma} = 1$ and $y_{1 \perp} = 0$, corresponding to noninteracting particles, and in the wire interactions lead to $K_{\sigma} \neq 1$ and $|y_{1 \perp}| > 0$ (see Sec.~\ref{sec:bosonization}).}
\label{fig:geometry}
\end{figure}

We consider a geometry where an interacting wire of finite size $L$ is connected to infinite, noninteracting leads on either side, as shown in Fig.~\ref{fig:geometry}. We model Fermi-liquid leads as one-dimensional noninteracting systems with the corresponding parameters of the bosonized Hamiltonian (see Sec.~\ref{sec:bosonization}). In the related experiments \cite{Krinner_spin_and_particle2016, Lebrat_band_and_correlated2018}, the particle reservoirs have attractive interactions but are at a finite temperature. If the temperature is above the spin gap, one can expect the effects of pairing to be negligible, so that noninteracting leads are a reasonable description. The leads enter the calculation of the conductance only in the zero-temperature limit of Sec.~\ref{sec:low_temperature}, whereas at high and intermediate temperatures, as defined in Sec.~\ref{sec:renormalization_group_gapped_and_gapless}, only the finite length of the wire plays a role and leads do not need to be considered explicitly.

\subsection{Spin conductance and spin drag}

The main quantities of interest in this study are the spin conductance, spin conductivity, and spin drag. Conductivity characterizes the linear current response to an external field -- an electric field in the case of charge conductivity and a magnetic field gradient in the case of spin conductivity. Conductivity, which measures the response in the thermodynamic limit, is usually related to the conductance $G$ of a wire of finite length $L$ as $G = \sigma/L$. In the case of two spin components, the linear response relation for conductance can be written in matrix form as
\begin{equation}
\begin{pmatrix}
I_{\uparrow} \\
I_{\downarrow}
\end{pmatrix}
=
\begin{pmatrix}
G_{\uparrow \uparrow}	&\Gamma  \\
\Gamma				 	&G_{\downarrow \downarrow}
\end{pmatrix}
\begin{pmatrix}
\Delta \mu_{\uparrow} 	\\
\Delta \mu_{\downarrow}
\end{pmatrix}.
\label{eq:up_down_conductance}
\end{equation}
The differences in chemical potential across the wire are denoted by 
$\Delta \mu_{\uparrow, \downarrow}$ for spin-up and spin-down particles, and the corresponding currents by $I_{\uparrow, \downarrow}$. We denote the cross-conductance, or spin drag, by~$\Gamma$. Physically, spin drag gives the proportionality of a spin-down current to a voltage on particles with spin up, or vice versa, whereas $G_{\uparrow \uparrow(\downarrow \downarrow)}$ gives the proportionality of a current of one spin species to a voltage on the same species. In a recent experiment, such spin-dependent chemical potential differences were realized as different spin population imbalances in two atom cloud reservoirs on either side of a quantum point contact \cite{Krinner_spin_and_particle2016}. 

Instead of the transport of spin-up and spin-down particles, we focus on the transport of the collective degrees of freedom, charge and spin, as discussed in the following section. Conductance can be defined for charge and spin in terms of the above quantities as $G_{\rho, \sigma} = (I_{\uparrow} \pm I_{\downarrow})/(\Delta\mu_{\uparrow} \pm \Delta\mu_{\downarrow})$. Here, the upper sign refers to charge, denoted by $\rho$, and the lower one to spin $\sigma$. The spin drag is now given by $\Gamma = (G_{\rho} - G_{\sigma})/2$, and we evaluate the charge and spin conductances using the bosonization description.

\subsection{Bosonization}
\label{sec:bosonization}

We consider a one-dimensional quantum wire of fermions with contact interactions, described by the continuum Hamiltonian
\begin{align}
\begin{split}
H = -\frac{\hbar^2}{2 m} &\sum_{s = \uparrow, \downarrow} \int \psi_{s}^{\dagger}(x) \frac{\partial^2}{\partial x^2} \psi_{s}(x) dx \\
&+ g_{1 \perp} \int \psi_{\downarrow}^{\dagger}(x) \psi_{\uparrow}^{\dagger}(x) \psi_{\uparrow}(x) \psi_{\downarrow}(x) dx.
\end{split}
\label{eq:gaudin-yang}
\end{align}
Here, $\hbar$ is the reduced Planck constant, $m$ the particle mass, $g_{1 \perp}$ the coupling constant for backscattering, and $\psi_s^{\dagger}$ ($\psi_s^{\phantom{\dagger}}$) the field operator for creating (destroying) a fermion of spin $s = \uparrow, \downarrow$. In the following, we set $\hbar = 1$, as for the Boltzmann constant $k_B = 1$. For the theoretical analysis, we work with the low-energy field theory model corresponding to Eq.~(\ref{eq:gaudin-yang}),
\begin{equation}
H = H_{\rho}^0 + H_{\sigma},
\label{eq:bosonization_hamiltonian}
\end{equation}
where $H_{\sigma} = H_{\sigma}^0 + H_{\sigma}'$. The Hamiltonian operators for the charge and spin are decoupled. The ``unperturbed'' Hamiltonian $H_{\nu}^0$, with $\nu = \rho, \sigma$, has the quadratic form
\begin{equation}
H_{\nu}^0 = \frac{1}{2 \pi} \int dx \left[ v_{\nu} K_{\nu} \left( \partial_x \theta_{\nu}(x, t) \right)^2 + \frac{v_{\nu}}{K_{\nu}}\left( \partial_x \phi_{\nu}(x, t) \right)^2 \right].
\label{eq:luttinger_liquid}
\end{equation}
This Hamiltonian provides an effective description of the low-energy properties of a wide class of microscopic models in one dimension.
Equation~(\ref{eq:luttinger_liquid}) is quadratic in the bosonic fields $\phi_{\nu}$ and $\theta_{\nu}$ and describes a Tomonaga-Luttinger liquid -- a critical system with correlations decaying as power laws. The exponents of the power laws are functions of the Luttinger parameters $K_{\rho}$ and $K_{\sigma}$, and the velocities $v_{\rho}$ and $v_{\sigma}$ of charge and spin excitations are in general different from each other. When the field-theory Hamiltonian is used as an effective description of a certain microscopic model, the parameters $K_{\nu}$ and $u_{\nu}$ are determined by the parameters of the original model, such as interactions. Here, we consider $K_{\nu}$ and $u_{\nu}$ more generally without restriction to a specific microscopic model, but point out the parameter regime relevant for an experiment with spin-rotation invariant contact interactions between fermions \cite{Krinner_spin_and_particle2016, Lebrat_band_and_correlated2018}. 

We consider a system where the charge degree of freedom is described by a TLL. In the case of a TLL wire connected to leads, it was previously shown that the conductance is given by the conductance quantum $e^2/h$, where $e$ is the elementary charge and $h$ the Planck constant, multiplied by the Luttinger parameter $K$ of the leads, $G = K e^2/h$ \cite{Safi_transport1995, Maslov_conductance1995, Ponomarenko_conductance1995}. Here, $K$ denotes the Luttinger parameter of spinless fermions. The same result was found for a spatially varying $K$ within the wire \cite{Thomale_minimal_model2011}. In the case of neutral atoms, as considered here, the particle conductance does not contain the electric charge and the conductance quantum is given by $1/h$. We consider noninteracting leads with $K_{\sigma} = K_{\rho} = 1$, in which case the charge conductance is simply given by the conductance quantum~$G_{\rho} = 1/h$. 

Whereas the charge degree of freedom is a TLL, the spin degree of freedom is described by the sine-Gordon model \cite{Giamarchi_one_dimension2003}. The spin Hamiltonian has the additional cosine term 
\begin{equation}
H_{\sigma}' = \frac{2 g_{1\perp}}{(2 \pi \alpha)^2} \int_{-\frac{L}{2}}^{\frac{L}{2}} dx \cos \left( 2 \sqrt{2} \phi_{\sigma}(x, t) \right),
\label{eq:sine-gordon}
\end{equation}
which arises from the backscattering of fermions with opposite spin within the interacting wire. The coupling $g_{1 \perp}$ denotes the backscattering amplitude, and the short-distance cutoff $\alpha$ is chosen as the distance between particles, or inverse density $\rho_0^{-1}$. The leads are described by Eq.~(\ref{eq:luttinger_liquid}) with $K_{\sigma} = 1$ and $v_{\sigma} = v_F$. For the spin conductance $G_{\sigma}$, no exact solution is available, and we use approximate analytic expressions to evaluate it in different parameter regimes. A cosine term similar to Eq.~(\ref{eq:sine-gordon}) would also arise in the charge sector in the presence of an umklapp term generated by a lattice potential at commensurate filling. Earlier theoretical studies have considered the effect of umklapp scattering on charge conductance in such systems~\cite{Ponomarenko_treshold_features1997, Ponomarenko_Mott_insulator1998, Ponomarenko_spin_gap_insulators2000, Ponomarenko_spin_transport_Mott2014, Morath_inhomogeneous_wires2016}.

While the TLL is gapless, the sine-Gordon model can have an energy gap for excitations. It is physically intuitive that spin excitations have a gap when the interactions are attractive since fermions with opposite spins form pairs. In terms of the Hamiltonian, a spin gap forms when energy is minimized by fixing the field $\phi_{\sigma}$ to one of the minima of the cosine. The cosine term can then be expanded around the minimum and approximated by a quadratic mass term. Whether the ground state of a given microscopic model is gapped or gapless is revealed by a renormalization group (RG) analysis.

\subsection{Renormalization group: gapped and gapless regime}
\label{sec:renormalization_group_gapped_and_gapless}

\begin{figure}[h]
\centering
\includegraphics[width=0.6\linewidth]{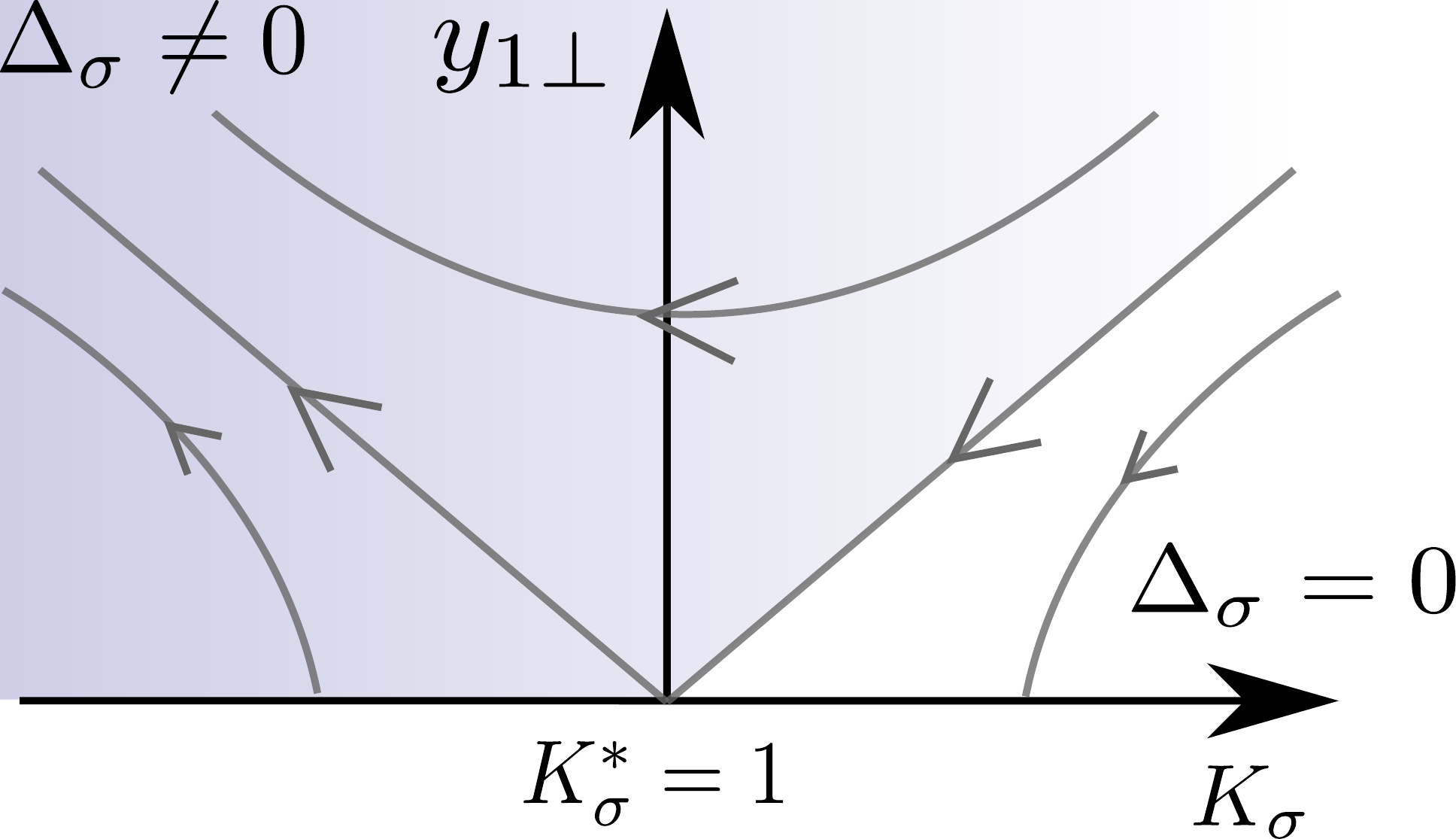}
\caption{The RG flow corresponding to Eqs.~(\ref{eq:RG_equations}). On the side of $K_{\sigma} > 1$, below the separatrix line $y_{1 \perp} \approx 2 K_{\sigma} - 2$, the coupling $y_{1 \perp}(l)$ approaches zero at $l \rightarrow \infty$ and the cosine term $H_{\sigma}'$ is irrelevant. Above the separatrix and for $K_{\sigma} < 1$, $y_{1 \perp}(l) \rightarrow \infty$ with increasing length scale. In this region, $H_{\sigma}'$ is relevant in the renormalization and there is a nonzero spin gap~$\Delta_{\sigma}$.}
\label{fig:RG_flow}
\end{figure}

For Hamiltonian $H_{\sigma}$, one has the RG equations (see e.g. Ref.~\onlinecite{Giamarchi_one_dimension2003})
\begin{align}
\begin{split}
\frac{d K_{\sigma}(l)}{d l} &= - \frac{1}{2} y_{1 \perp}(l)^2 K_{\sigma}(l)^2, \\
\frac{d y_{1 \perp}(l)}{d l} &= [2 - 2 K_{\sigma}(l)] y_{1 \perp}(l),
\end{split}
\label{eq:RG_equations}
\end{align}
where $y_{1 \perp} = g_{1 \perp}/(\pi v_{\sigma})$. 
Here, $l$ is the length scale changed in renormalization. The RG equations show that for an infinite wire, the sine-Gordon model has a gapped and a gapless parameter regime as shown by the RG flow in Fig.~\ref{fig:RG_flow}.
The separatrix $y_{1 \perp}(0) = 2 \left|K_{\sigma}^2(0) - 1 \right|/ \left|K_{\sigma}^2(0) + 1 \right|$ corresponds to interactions that are spin-rotation invariant [see for example Eq.~(2.105) in Ref.~\onlinecite{Giamarchi_one_dimension2003}]. At $K_{\sigma}$ close to 1, one can approximate $y_{1 \perp} \approx 2 K_{\sigma} - 2$. The contact interactions between fermions with opposite spin, as realized in the transport experiments of Refs.~\onlinecite{Krinner_spin_and_particle2016, Lebrat_band_and_correlated2018, Corman_dissipative_point_contact2019}, would fall on this line. Since the RG equations are the same for $y_{1 \perp} > 0$ and $y_{1 \perp} < 0$, the RG flow is symmetric with respect to the $y_{1 \perp} = 0$ axis. We can therefore consider only the $y_{1 \perp} > 0$ half-plane. The sign of the interaction is encoded in the value of $K_{\sigma}$: $K_{\sigma} < 1$ corresponds to attractive and $K_{\sigma} > 1$ to repulsive interactions.

In the following, we use the RG equations to analyze the spin conductance of a finite-length wire. We consider different parameter regimes in terms of the wire length $L$, thermal length $L_T = v_{\sigma}/T$, and superfluid coherence length $L_{\Delta} = v_{\sigma}/\Delta_{\sigma}$. The discussion of the different parameter regimes is structured into sections for the high ($L_T \ll L, L_{\Delta}$), intermediate ($L \ll L_T \ll L_{\Delta}$ and $L_{\Delta} \ll L_T \ll L$), and low temperature ($L, L_{\Delta} \ll L_T$). In the intermediate and low-temperature regions, we additionally make a distinction between $L \ll L_{\Delta}$ and $L_{\Delta} \ll L$. 

The shortest of the lengths $L$, $L_{\Delta}$, and $L_T$ acts as a limiting length scale in the renormalization: the parameters $y_{1 \perp}(l)$, $K_{\sigma}(l)$, and the cutoff $\alpha(l) = \alpha(0) e^{l}$ are renormalized up to a length scale $l^*$ at which $\alpha(l^*)$ reaches one of these lengths. By renormalization, one finds the parameters which describe the low-energy behavior of the model; after finding these new parameters, one still usually has to calculate the desired quantities in the new model by other means. Depending on the parameter regime, we use perturbation theory, an instanton approach, or a two- or three-step RG procedure to calculate the spin conductance using the renormalized parameters.
When only repulsive interactions are considered, the cosine term of Eq.~(\ref{eq:sine-gordon}) is often left out since it is irrelevant in the asymptotic limit. Instead, other scattering sources, beyond the TLL approximation, have been considered for the charge and spin conductance~\cite{Matveev_Wigner-crystal2004, Matveev_low_electron_density2004, Meyer_Wigner_crystal2008}. 
Here, we focus on spin transport and take the cosine term into account explicitly. In the attractive case, it is relevant in renormalization, but it also gives contributions in the repulsive case due to the finite length of the wire.

\subsection{Spin conductance in different temperature regimes}

\begin{figure}[h]
\centering
\includegraphics[width=\linewidth]{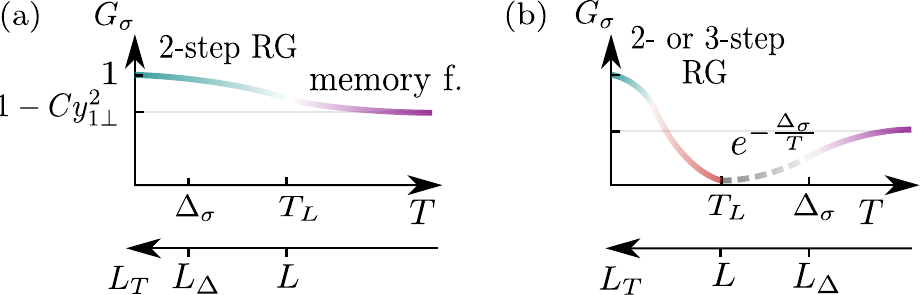}
\caption{A sketch of the spin conductance as a function of temperature for the different orders of energy scales $\Delta_{\sigma}$ and $T_L = v_{\sigma}/L$. Panel (a) corresponds to a weak coupling $y_{1 \perp}$ with $\Delta_{\sigma} \ll T_L$ ($L \ll L_{\Delta}$) and (b) to strong coupling with $T_L \ll \Delta_{\sigma}$ ($L_{\Delta} \ll L$). In both cases, the spin conductance reaches the value 1 (the conductance quantum) at $T = 0$, whereas at high temperature, $G_{\sigma}$ has a correction proportional to $y_{1 \perp}^2$. In panel (b), $y_{1 \perp}$ is larger and the value of $G_{\sigma}$ in the high-temperature limit is lower.  The different methods used for computing the temperature dependence in the various temperature regimes are marked with different colors or line styles (see text). The limits of validity of each method are depicted with shaded lines.}
\label{fig:sketch}
\end{figure}

The finite-length wire has conducting and insulating phases depending on the relative magnitudes of the length scales $L$, $L_{\Delta} = v_{\sigma}/\Delta_{\sigma}$, and $L_T = v_{\sigma}/T$. The temperature dependence and, correspondingly, thermal-length dependence of the spin conductance is illustrated schematically in Fig.~\ref{fig:sketch}. Panel (a) shows the case of a weak coupling $y_{1 \perp}$ with $\Delta_{\sigma} \ll T_L$ ($L \ll L_{\Delta}$) and panel (b) corresponds to strong coupling with $T_L \ll \Delta_{\sigma}$ ($L_{\Delta} \ll L$). Whether the coupling is strong or weak by this criterion depends both on the interaction and on the wire length: for a finite length of the wire, the system can be in the weak-coupling regime both for repulsive and sufficiently small attractive interactions.

We employ different methods to compute the temperature dependence of the spin conductance in the different temperature regimes, as indicated by the different line styles and colors in Fig.~\ref{fig:sketch}. When temperature is the highest energy scale, $T \gg T_L, \Delta_{\sigma}$, one can use the memory-function approach discussed in Sec.~\ref{sec:high_temperature}, independent of the relative values of $T_L$ and $\Delta_{\sigma}$.  
The memory-function calculation shows that the spin conductance has the dependence ${G_{\sigma} - 1 \propto -y_{1 \perp}^2 (L/\alpha) (T/\Lambda)^{4 K_{\sigma} - 3}}$ on temperature and coupling. 
This expression shows that if $K_{\sigma} > 3/4$, there is a decrease of conductance with temperature in the high-temperature regime, whereas for $K_{\sigma} < 3/4$, conductance grows with temperature. Both cases can occur for both weak and strong coupling, as discussed in Sec.~\ref{sec:memory_function}, but $K_{\sigma} > 3/4$ is more representative of the weak-coupling regime and $K_{\sigma} < 3/4$ of the strong-coupling regime in the case of spin-rotation invariant interactions. The drawings in Fig.~\ref{fig:sketch} represent these typical cases.
The memory-function expression is valid up to temperatures on the order of the high-energy cutoff $\Lambda = v_{\sigma}/\alpha$, where $G_{\sigma}$ reaches the value $G_{\sigma} = 1 - Cy_{1 \perp}^2$ with the constant $C$ of order 1. The lower limit of validity of the memory-function approach is when the thermal length reaches the smaller of either the wire length or the superfluid coherence length, below which one can expect a different behavior.

When $T < T_L$, we use the renormalization-group procedures outlined in Sec.~\ref{sec:low_temperature}.
In the case of weak coupling and low temperature $L \ll L_{\Delta}, L_T$, shown in Fig.~\ref{fig:sketch}(a), the problem reduces to a single interacting point in an otherwise noninteracting wire. As discussed in Sec.~\ref{sec:weak_coupling}, an RG analysis shows that the zero-dimensional interacting system within a noninteracting wire is irrelevant, and the wire is perfectly conducting at the low-energy limit, $G_{\sigma}(T = 0) = 1$. 

\begin{figure}[h]
\centering
\includegraphics[width=0.95\linewidth]{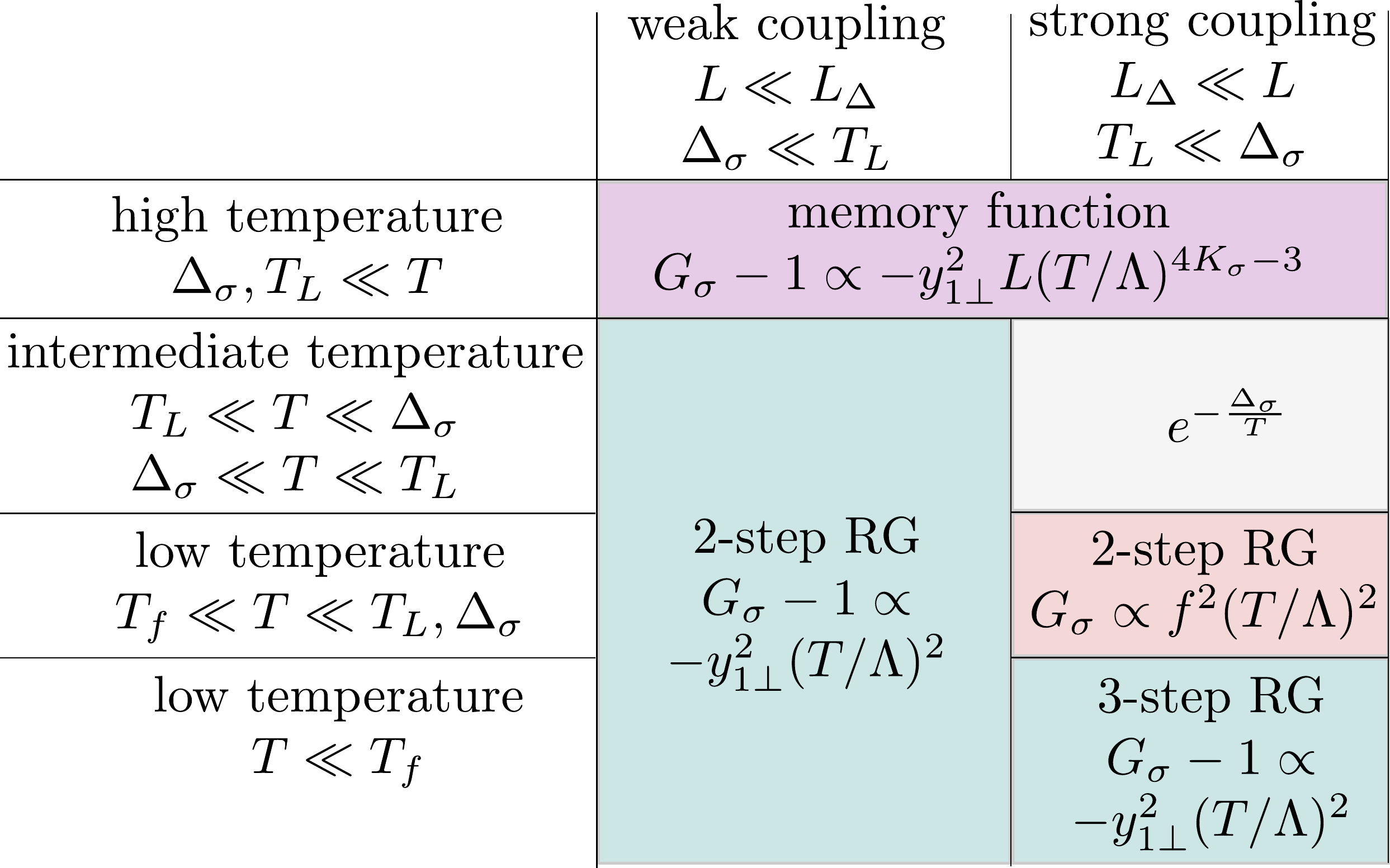}
\caption{A summary of the different parameter regimes. The relevant techniques and temperature dependence of the spin conductance found in these regimes correspond to the schematic drawing of Fig.~\ref{fig:sketch}. The temperature $T_f$ is related to the three-step renormalization group procedure detailed in Sec.~\ref{sec:strong_coupling}.}
\label{fig:regimes}
\end{figure}

In the case of strong coupling, there is an intermediate-temperature regime with an exponential dependence of conductance on temperature when the temperature reduces below the spin gap, marked with a dashed line. This temperature regime is discussed in Sec.~\ref{sec:intermediate_temperature}. As detailed in Sec.~\ref{sec:strong_coupling}, for a finite-length wire with a finite energy scale $T_L$, renormalization group arguments again show that perfect spin conductance is recovered in the zero-temperature limit. This is in contrast with the vanishing spin conductivity in the gapped phase (see Appendix~\ref{app:conductivity}). The different line colors for $T < T_L$ indicate a renormalization procedure where either the coupling $y_{1 \perp}$ (green) or the fugacity $f$ (orange), as defined in Sec.~\ref{sec:strong_coupling}, is used as a perturbative parameter. The different parameter regimes and the relevant methods and results in these regimes are summarized in Fig.~\ref{fig:regimes}. 

While a nonmonotonic temperature dependence can occur also for weak coupling, the nonmonotonic behavior is more distinct for strong coupling, where the spin conductance reaches a lower (nonzero) minimum at $T_L$.
This regime is therefore more interesting from an experimental point of view. 
In Sec.~\ref{sec:prediction}, we construct the temperature dependence of $G_{\sigma}$ for experimentally relevant parameters, using the different methods described in Secs.~\ref{sec:high_temperature}--\ref{sec:low_temperature}.

\section{Spin conductance at high temperature $L_T \ll L, L_{\Delta}$}
\label{sec:high_temperature}

When the temperature is larger than all other energy scales, $L_T \ll L, L_{\Delta}$, we use the thermal length scale $L_T = v_{\sigma}/T$ as a limiting criterion in the renormalization: the parameters are renormalized up to a length scale $l^*$ for which $\alpha(l^*) = L_T$. The renormalized parameters are used for calculating the spin conductance perturbatively by the memory-function formalism. The approach to calculating the conductance in the high-temperature regime is the same for weak and strong coupling ($L \ll L_{\Delta}$ and $L_{\Delta} \ll L$).

\subsection{Renormalization of parameters}
\label{sec:renormalization_of_parameters}

\begin{figure}[h]
\includegraphics[width=\linewidth]{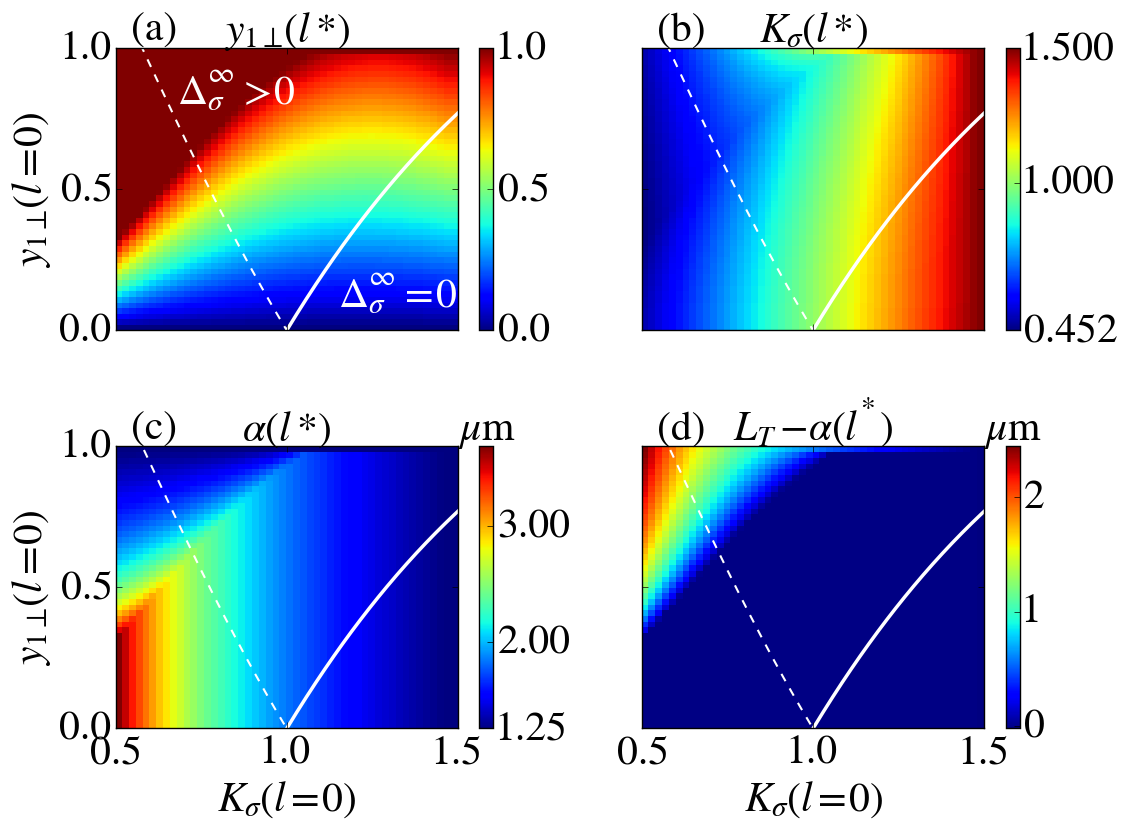}
\caption{The renormalized parameters (a) $y_{1 \perp}(l^*)$ and (b) $K_{\sigma}(l^*)$, (c) the cutoff $\alpha(l^*) = \alpha_0 e^{l^*}$, and (d) the cutoff subtracted from the thermal length $L_T - \alpha(l^*)$, as functions of $K_{\sigma}(l = 0)$ and $y_{1 \perp}(l = 0)$. The parameters have been evolved according to Eqs.~(\ref{eq:RG_equations}) up to the length scale $l^*$ at which either $\alpha(l^*) \approx L_T$ or $y_{1 \perp}(l^*) \approx 1$. The white lines show the separatrix which corresponds to spin-rotation invariant interactions.  Along the solid line, the parameters flow towards the fixed point $(K_{\sigma} = 1, y_{1 \perp} = 0)$ and along the dashed line, towards $y_{1 \perp} \rightarrow \infty$. The region to the left of the solid line has a finite spin gap in the thermodynamic limit, $\Delta_{\sigma}^{\infty} > 0$, and the region to the right is gapless in the thermodynamic limit, $\Delta_{\sigma}^{\infty} = 0$ (see Fig.~\ref{fig:RG_flow}). We have used here $\alpha(l = 0) = \rho_0^{-1} = 1.25 \mu \text{m}$ with $\rho_0 = 0.8/\mu \text{m}$. Since $L_T < L = 5.5 \mu$m, the wire length does not enter here as a limiting length in the renormalization.}
\label{fig:renormalized_parameters_LT}
\end{figure}

When the thermal length is smaller than other length scales, the wire can be viewed as a series of incoherent blocks of length $L_T$. Temperature can then be incorporated in the RG procedure by renormalizing the parameters $K_{\sigma}$ and $y_{1 \perp}$ up to the length scale $l^*$ at which $\alpha(l^*) \sim L_T$. To illustrate the evolution of the parameters with increasing length scale, Figure~\ref{fig:renormalized_parameters_LT} shows the renormalized parameters as functions of $K_{\sigma}(l = 0)$ and $y_{1 \perp}(l = 0)$ at the original length scale at a fixed temperature $T = 55$~nK, corresponding to the experimental conditions of Ref.~\onlinecite{Krinner_spin_and_particle2016}. In the figure, we have renormalized the parameters up to the length scale $l^*$ at which either i) $\alpha(l^*) \sim L_T$ or ii) $y_{1 \perp}(l^*) \sim 1$, corresponding to $\alpha(l^*) \sim L_{\Delta}$. The latter condition is due to the fact that the RG equations are perturbative in $y_{1 \perp}$ and not valid beyond $y_{1 \perp}(l^*) \simeq 1$. The thermal length is calculated as $L_T = v_{\sigma}/T$, where the spin velocity remains essentially unchanged in renormalization and is given by $v_{\sigma} = v_F/K_{\sigma}(0)$ for a Galilean invariant system. 

As shown in Fig.~\ref{fig:renormalized_parameters_LT}(d), $\alpha(l^*) = L_T$ and $y_{1 \perp}(l^*) < 1$ in the majority of the diagram whereas in the upper left corner, $\alpha(l^*) < L_T$ and $y_{1 \perp}(l^*) = 1$. The memory-function approach discussed in the following section is only valid for the region with $y_{1 \perp}(l^*) < 1$. The white lines show the separatrix 
\begin{equation}
y_{1 \perp}(0) = 2 \left| \frac{K_{\sigma}^2(0) - 1}{K_{\sigma}^2(0) + 1} \right|
\label{eq:separatrix}
\end{equation}
which corresponds to spin-rotation invariant interactions. The solid white line separates the regions with a finite spin gap in the thermodynamic limit $\Delta_{\sigma}^{\infty} > 0$, where $y_{1 \perp}$ flows to strong coupling, and the region that is gapless in the thermodynamic limit $\Delta_{\sigma}^{\infty} = 0$. Due to the finite size of the wire, the region where $y_{1 \perp}(l^*) \approx 1$ does not exactly match the one for which $\Delta_{\sigma}^{\infty} > 0$. Along the solid line, the parameters flow towards the fixed point $(K_{\sigma} = 1, y_{1 \perp} = 0)$ and along the dashed line, towards $y_{1 \perp} \rightarrow \infty$.

\subsection{Memory function}
\label{sec:memory_function}

To calculate the spin conductance in the high-temperature case, we couple RG with a memory-function calculation. When the thermal length is smaller than other length scales, one can neglect the finite length of the wire and calculate the spin conductance from the spin conductivity, $G_{\sigma} = \sigma_{\sigma}/L$. The spin conductivity has an expression in terms of the spin current-current correlation, which can be evaluated using bosonization. When the backscattering term $H_{\sigma}'$ is present, this correlation function cannot be obtained exactly but one can calculate it perturbatively when the perturbation expansion in $y_{1 \perp}$ converges. This is the case when $y_{1 \perp}$ flows to weak coupling in the renormalization, or flows to strong coupling but the renormalization is stopped at $\alpha(l^*) \sim L_T$ before $y_{1 \perp}(l) \sim 1$. We obtain the conductivity by calculating the current-current correlation in a second-order perturbation expansion \cite{Giamarchi_umklapp_resistivity1991}. 

The spin conductivity can be written in terms of the memory function, as shown in Appendix~\ref{app:memory_function}. One can obtain the dependence on $y_{1 \perp}$ and $T$ in the form
\begin{equation*}
\sigma_{\sigma}(T) \propto \frac{\alpha}{y_{1 \perp}^2} \left( \frac{T}{\Lambda} \right)^{3 - 4 K_{\sigma}},
\end{equation*}
where $\Lambda = v_{\sigma}/\alpha$ is a high-energy cutoff. The functional form by itself is however only accurate when $K_{\sigma}$ is unchanged in renormalization. In the situation relevant for the experiment, $K_{\sigma}$ is on the separatrix and is renormalized according to Eq.~(\ref{eq:RG_equations}). Therefore, it is necessary to couple the memory-function expression with the renormalization of the parameters. The conductance of the wire can be calculated by adding in series the resistance of the wire and the contact resistance. The resistance of the wire is $R_{\text{wire}} = \rho L$ and the contacts have a resistance equal to the inverse conductance quantum, $R_{\text{contact}} = 1$. The spin conductance can be written as
\begin{equation}
G_{\sigma} = \frac{1}{R_{\text{wire}} + R_{\text{contact}}} = \frac{1}{R_{\text{wire}} + 1} \approx 1  - R_{\text{wire}},
\label{eq:resistances}
\end{equation}
where the approximation is valid when $R_{\text{wire}}$ is small. Using the expression $\rho(T) \propto y_{1 \perp}^2 (T/\Lambda)^{4 K_{\sigma} - 3}/\alpha$ for the resistivity gives the dependence
\begin{equation}
G_{\sigma}(T) - 1 \propto -y_{1 \perp}^2 \frac{L}{\alpha} \left( \frac{T}{\Lambda} \right)^{4 K_{\sigma} - 3}
\label{eq:proportionality}
\end{equation}
on the coupling, temperature, wire length, and cutoff. 

As noted above, Eq.~(\ref{eq:proportionality}) involves certain subtleties concerning the renormalization of the parameters $y_{1 \perp}$, $K_{\sigma}$, and $\alpha$. In a situation where $y_{1 \perp} \approx 0$ and $K_{\sigma}$ is away from the separatrix, one can approximate the RG flow to be vertical: $d K_{\sigma}/dl = 0$. In this case, the relation~(\ref{eq:proportionality}) is invariant with changing length scale and gives the correct temperature dependence of the spin conductance for both $y_{1 \perp}(l = 0), \alpha(l = 0)$ and $y_{1 \perp}(l^*), \alpha(l^*) = L_T$. The exponent shows that the spin conductance has a different behavior depending on whether $K_{\sigma} < 3/4$ or $K_{\sigma} > 3/4$. For $K_{\sigma} < 3/4$, the exponent is negative and $G_{\sigma}$ decreases with decreasing temperature. For $K_{\sigma} > 3/4$ on the other hand, the exponent is positive and the correction to $G_{\sigma} = 1$ decreases with temperature, giving therefore an increase of $G_{\sigma}$ with lowering temperature. 
As the renormalized value of $K_{\sigma}$ depends on the interactions, particle density, and temperature, both cases $K_{\sigma} < 3/4$ and $K_{\sigma} > 3/4$ can occur for both weak and strong coupling. For spin-rotation invariant interactions, the case with $K_{\sigma} < 3/4$ mostly coincides with the strong-coupling regime.

The temperature dependence of Eq.~(\ref{eq:proportionality}) applies to temperatures between the high-energy cutoff $\Lambda$ and the spin gap $\Delta_{\sigma}$, below which the memory-function approach is not valid. The behavior of the spin conductance at $T < \Delta_{\sigma}$ is discussed in Secs.~\ref{sec:intermediate_temperature} and~\ref{sec:low_temperature}. When $y_{1 \perp}$ is not so small, or if $K_{\sigma}$ is fixed to the separatrix, so that $d K_{\sigma}/dl \neq 0$, Eq.~(\ref{eq:proportionality}) is approximate and the accuracy can be improved by renormalizing the parameters. In this case, the temperature dependence of the correction to $G_{\sigma} = 1$ is nontrivial and can be obtained numerically. Section~\ref{sec:prediction} shows the spin conductance as a function of temperature evaluated numerically for experimentally relevant parameters.

\subsection{Comparison to experimental data}
\label{sec:memory_function_comparison}

In order to compare the memory-function result to experimental data~\cite{Krinner_spin_and_particle2016}, we evaluate the spin conductance of Eq.~(\ref{eq:resistances}) for experimentally relevant parameters. The resistance $R_{\text{wire}} = L/\sigma_{\sigma}$ is calculated by using the memory-function expression~(\ref{eq:conductivity_at_omega0}) of Appendix~\ref{app:memory_function} for the spin conductivity $\sigma_{\sigma}$. The coupling $y_{1 \perp}$ is determined by the scattering length~$a$ and the particle density~$\rho_0$, as detailed in~Appendix~\ref{app:experimental_parameters}. The experimental parameters used in this calculation are also listed in Appendix~\ref{app:experimental_parameters}. The s-wave scattering between fermions in different spin states, present in the experiment~\cite{Krinner_spin_and_particle2016}, can be modeled by contact interactions which are spin-rotation invariant. We therefore consider RG flow along the separatrix, with $K_{\sigma}$ fixed by $y_{1 \perp}$ as in Eq.~(\ref{eq:separatrix}). Note that the RG procedure and the temperature dependence of the spin conductivity found here is valid beyond this specific type of interaction -- they are valid for any interaction which decays as $1/x^{\gamma}$ with $\gamma > 1$, and may or may not be spin-rotation invariant~\cite{Giamarchi_one_dimension2003}. 

\begin{figure}[h]
\includegraphics[width=\linewidth]{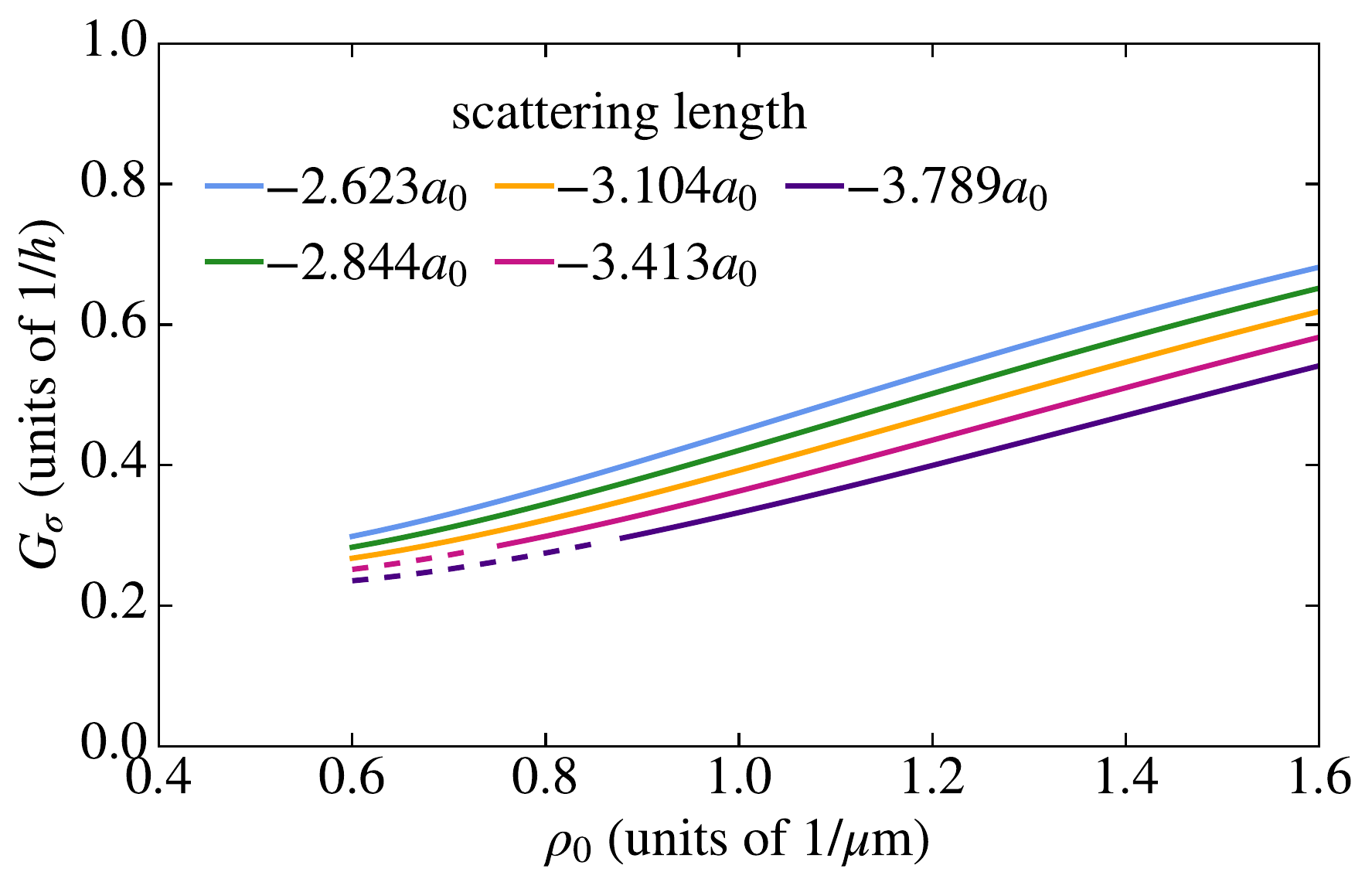}
\caption{The spin conductance as a function of $\rho_0$ for the same values of the scattering length as in Fig.~\ref{fig:experimental}, calculated using Eqs.~(\ref{eq:resistances}), (\ref{eq:memory_function}), and (\ref{eq:conductivity_at_omega0}) with renormalized parameters. Here, {$a_0 = 5.25\cdot 10^{-11}$ m} is the Bohr radius and the wire length is $L = 5.5 \mu$m~\cite{Krinner_spin_and_particle2016}. The dashed line indicates a region where $T < \Delta_{\sigma}^{\text{ex}}$ (see text) and the memory-function expression is not reliable.}
\label{fig:memory_function}
\end{figure}

\begin{figure}[h]
\includegraphics[width=\linewidth]{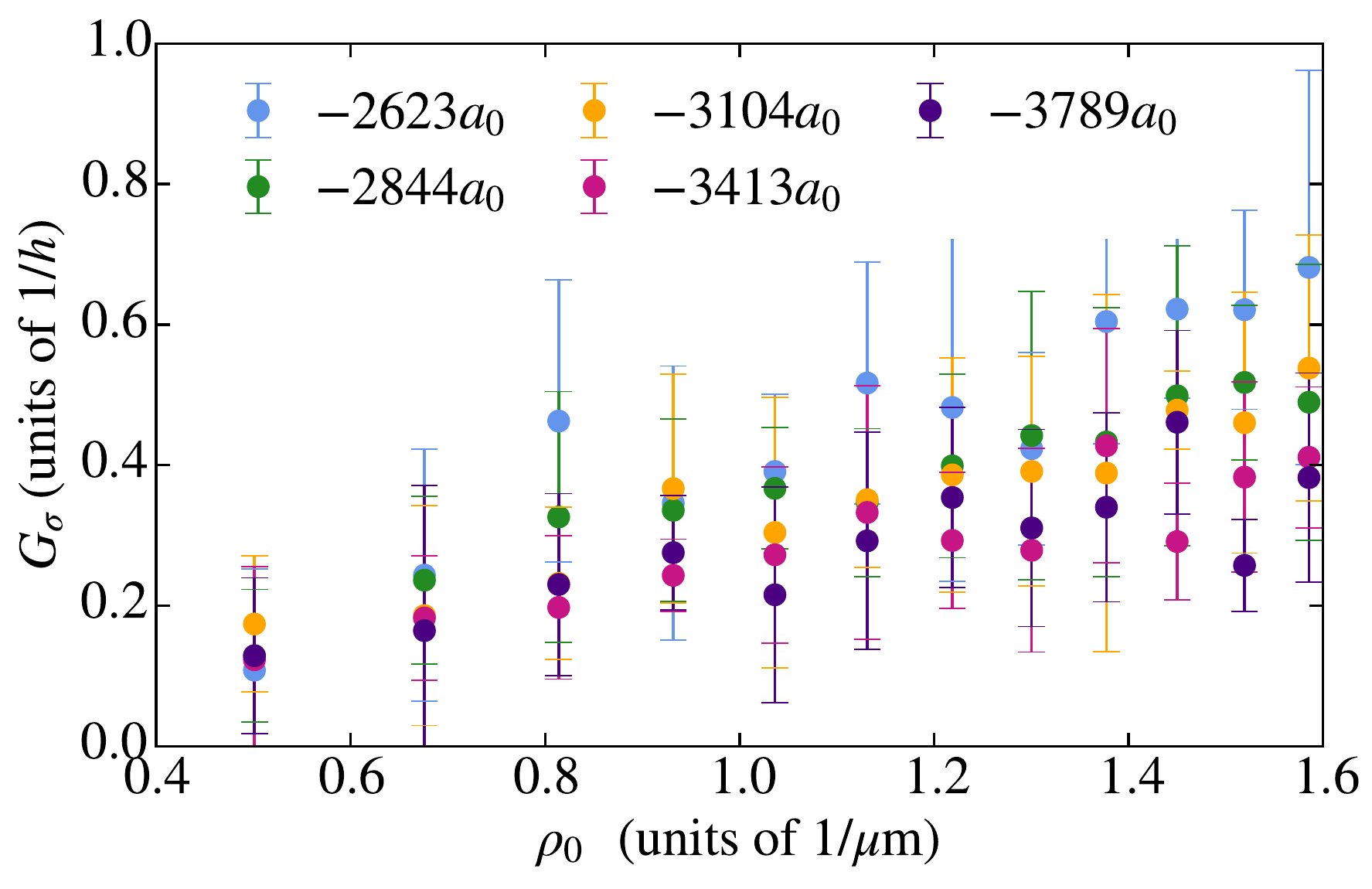}
\caption{The spin conductance measured as a function of the particle density in a quantum point contact setup~\cite{Krinner_spin_and_particle2016}, for various values of the scattering length.}
\label{fig:experimental}
\end{figure}

Figure~\ref{fig:memory_function} shows the memory-function result, which can be compared to the experimental data in Fig.~\ref{fig:experimental}. The data is part of the measurements done in Ref.~\onlinecite{Krinner_spin_and_particle2016}, where spin conductance is measured through a narrow channel between two particle cloud reservoirs. When only one transversal mode in the channel is occupied, we can consider it a one-dimensional system.
Due to the inhomogeneous potential profile in the experiment, there are higher-density regions at the entrance and exit of the channel. For strong interaction (large negative scattering length) and large $\rho_0$, these high-density regions transition from the normal state to superfluid, which leads to a nonmonotonic dependence of the spin conductance on particle density. The consequences of superfluidity in the leads on particle and spin conductance were discussed in recent theoretical work~\cite{Kanasz-Nagy_anomalous_conductances2016, Uchino_anomalous_transport2017}.

Since our model only describes the regime with leads in the normal state, we only show data extending up to $\rho_0 = 1.6/\mu$m, where the normal-to-superfluid transition is estimated to occur for the scattering length $a \approx -3500 a_0$. This density corresponds to the gate potential $V_g \approx 0.8 \mu$K in Fig.~2 in Ref.~\onlinecite{Krinner_spin_and_particle2016}. For small particle densities, another constraint is set by the limits of validity of the field-theory approach. We use the interparticle separation $\rho_0^{-1}$ as the short-distance cutoff $\alpha(l = 0)$, and therefore the comparison is restricted to particle densities for which the interparticle separation is smaller than the thermal length $L_T = v_{\sigma}/T$. For $T = 55$ nK, the comparison is limited to $\rho_0 \gtrsim 0.6/\mu$m. 
Additionally, the memory-function expression is only valid for temperatures above the spin gap, with a crossover region around $T \approx \Delta_{\sigma}$.
We have used a dashed line for the part where $T < \Delta_{\sigma}^{\text{ex}}$ in Fig.~\ref{fig:memory_function}, indicating that in this region the result may be unreliable. Here, $\Delta_{\sigma}^{\text{ex}}$ is the exact expression of the spin gap in the Gaudin-Yang model~\cite{Fuchs_exactly_solvable2004} (see Sec.~\ref{sec:prediction} and Appendix~\ref{app:experimental_parameters}).

Figures~\ref{fig:memory_function} and~\ref{fig:experimental} show that the memory-function calculation gives the same trend of increasing conductance with increasing $\rho_0$ as in the experimental data. In both figures, the spin conductance is lower for larger negative scattering lengths for which the spin-up and spin-down fermions are more strongly paired. Since the number of particles should in general be large for a field-theory description to be applicable, it is not obvious that it is accurate here. Figure~\ref{fig:memory_function} corresponds to between 3.3 and 8.8 particles within the wire. We however find a good qualitative agreement between the memory-function calculation and the experimental data, especially for the smallest values of $|a|$ and for low densities in the case of the largest $|a|$. This is consistent with the fact that we only consider leads in the normal state.

As mentioned above, we effectively view the wire as a series of incoherent blocks of length $L_T$. Such a consideration would be valid for a wire which is coupled to an environment that destroys phase coherence beyond the length scale~$L_T \ll L$. One therefore assumes that there are no conservation laws which would lead to a finite steady-state current.  In one dimension, or in an isolated system, this assumption is not necessarily valid, for example if the system is integrable. In the absence of phonons or other dissipative processes, integrable systems can have a current that saturates at a nonzero value instead of decaying to zero as a function of time. Therefore, the memory-function solution could be inaccurate in systems which do not have dissipation mechanisms, such as cold-atom experiments. Nevertheless, the good qualitative agreement between the memory-function result and the experimental data seems to validate the hypothesis that coherence is lost beyond the thermal length~$L_T$. It remains to be understood what the mechanism for loss of coherence is in the cold-atom experiment.

\section{Intermediate temperature $L \ll L_T \ll L_{\Delta}$ and $L_{\Delta} \ll L_T \ll L$}
\label{sec:intermediate_temperature}

In the case of strong coupling and a temperature below the spin gap, in the regime where $L_{\Delta} \ll L_T \ll L$, one can expect a very small spin conductance. The conductance has indeed an exponential dependence on the temperature and the spin gap, $G_{\sigma} \propto e^{-\Delta_{\sigma}/T}$. One can understand this in terms of tunneling events across the gap called instantons~\cite{Rice_new_excitation1976, Nattermann_variable-range2003}. At even lower temperature, when the thermal length exceeds the wire length, one can expect yet a different behavior as discussed in Sec.~\ref{sec:low_temperature}. In the case of weak coupling $L \ll L_{\Delta}$, the interacting wire reduces to a backscattering term at $x = 0$ which is irrelevant in renormalization. Therefore, the conductance has the same temperature dependence at $L \ll L_{\Delta} \ll L_T$ and $L \ll L_T \ll L_{\Delta}$, as indicated by the green line in Fig.~\ref{fig:sketch}(a).

\section{Low temperature $L, L_{\Delta} \ll L_T$}
\label{sec:low_temperature}

When the thermal length is the largest length scale, the smaller of $L$ and $L_{\Delta}$ enters as the limiting length in the renormalization procedure. 
We analyze the spin conductance at low temperature by renormalizing the parameters up to a length scale $l^*$ at which one of two criteria is reached: i) either $\alpha(l^*) \sim L$ 
or ii) $y_{1 \perp}(l^*) \sim 1$. In both cases, we find the spin conductance by relating the finite-length wire to a zero-dimensional system with backscattering. One can then employ RG equations that are different from those of the original problem. In the case of criterion i), we use a two-step RG procedure, meaning two different sets of RG equations applied consecutively, whereas for criterion ii), there are either two or three steps, as detailed in Sec.~\ref{sec:strong_coupling}.
Appendix \ref{app:renormalization_of_parameters} shows how the renormalized parameters $K_{\sigma}(l^*)$ and $y_{1 \perp}(l^*)$ depend on $K_{\sigma}(l = 0)$ and $y_{1 \perp}(l = 0)$.

\subsection{Weak coupling $L \ll L_{\Delta}(l^*) \ll L_T$: two-step renormalization group}
\label{sec:weak_coupling}

Weak coupling corresponds to criterion i) where the cutoff reaches the length of the wire before the coupling becomes non-perturbative. This parameter regime therefore includes both repulsive and weakly attractive interactions for which $y_{1 \perp}(l^*) < 1$, as shown in Fig.~\ref{fig:renormalized_parameters}. In this case, we identify the problem of a finite-length wire connected to noninteracting leads with that of a zero-dimensional system where particles backscatter, embedded in a noninteracting wire. 

\begin{figure}
\centering
\includegraphics[width=0.5\linewidth]{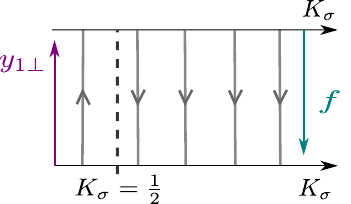}
\caption{The RG flow of $y_{1 \perp}$ and $K_{\sigma}$ corresponding to Eqs.~(\ref{eq:RG_point_impurity}) together with the flow of the fugacity $f$ defined in Sec.~\ref{sec:strong_coupling}. As $K_{\sigma}$ does not change in renormalization, the flow is vertical. For $K_{\sigma} > 1/2$, the coupling $y_{1 \perp}(l)$ approaches zero at $l \rightarrow \infty$, whereas for $K_{\sigma} < 1/2$, the coupling approaches infinity. The flow is the opposite for $f$, as seen from Eq.~(\ref{eq:RG_fugacity}). Note that the critical value $K_{\sigma} = 1/2$ is different than in the case of an impurity in a TLL~\cite{Kane_transport1992, Kane_transmission1992}. Here, $K_{\sigma} = 1$ in the leads means that $f$ grows in renormalization while $y_{1 \perp}$ flows to zero.}
\label{fig:RG_flow_backscattering_fugacity}
\end{figure}

The cutoff $\alpha$, which here is fixed as the inter-particle separation, can be thought of as the shortest length scale at which $\phi_{\sigma}$ varies. As $\alpha$ reaches $L$ in renormalization, the integral in Eq.~(\ref{eq:sine-gordon}) is taken over an interval of length $\alpha$ within which $\phi_{\sigma}(x)$ is constant. We can therefore replace $\cos \left(2 \sqrt{2} \phi_{\sigma}(x, t) \right)$ by $\cos \left(2 \sqrt{2} \phi_{\sigma}(x = 0, t) \right)$, which leads to the term
\begin{align}
H_{\sigma}' =  \frac{2 g_{1 \perp}}{4 \pi^2 L} \cos \left(2 \sqrt{2} \phi_{\sigma}(x = 0, t) \right).
\label{eq:point_impurity}
\end{align}
Here, $\alpha^2$ in the denominator has been replaced with $L^2$ and a factor of $L$ in the numerator is given by the integration in Eq.~(\ref{eq:sine-gordon}). In this case, one has a different set of RG equations,
\begin{align}
\begin{split}
\frac{d K_{\sigma}(l)}{d l} &= 0, \\
\frac{d y_{1 \perp}(l)}{d l} &= [1 - 2 K_{\sigma}(l)] y_{1 \perp}(l) = -y_{1 \perp}(l).
\end{split}
\label{eq:RG_point_impurity}
\end{align}
The second equation gives the expression 
\begin{equation*}
y_{1 \perp} = y_{1 \perp}(0) e^{[1 - 2 K_{\sigma}] l}
\end{equation*}
for the coupling. The RG flow corresponding to Eqs.~(\ref{eq:RG_point_impurity}) is illustrated in Fig.~\ref{fig:RG_flow_backscattering_fugacity}, where one can see that $y_{1 \perp}$ flows to infinity whenever $K_{\sigma} < 1/2$ and to zero when $K_{\sigma} > 1/2$. There is a fixed line at $K_{\sigma} = 1/2$ where $y_{1 \perp}$ does not change in renormalization. As $K_{\sigma}$ now refers to the value in the leads, $K_{\sigma} = 1$, the coupling $y_{1 \perp}(l) = y_{1 \perp}(0) e^{-l}$ approaches zero at increasing length scale. Backscattering at $x = 0$ is therefore irrelevant at low energies and one obtains a model of free fermions with $G_{\sigma} = 1$ at $T = 0$.
One can find the temperature-dependent correction to the zero-temperature value of the spin conductance perturbatively. Similar to the weak-coupling solution in Ref.~\onlinecite{Kane_transmission1992}, we obtain the scaling
\begin{equation}
G_{\sigma} - 1 \propto - y_{1 \perp}^2 \left(\frac{T}{\Lambda} \right)^{4 K_{\sigma} - 2} = - y_{1 \perp}^2 \left(\frac{T}{\Lambda} \right)^2
\label{eq:perturbative_y1}
\end{equation}
when $K_{\sigma} = 1$ in the leads.

A similar problem of a point-like impurity potential in a Luttinger liquid was considered earlier by Kane and Fisher~\cite{Kane_transport1992, Kane_transmission1992}. The earlier study considered spinless fermions, for which the backscattering from an impurity potential is described by the term $\psi_R^\dagger \psi_L^{\phantom{\dagger}} + \text{H.c.}$, where $R$ denotes right- and $L$ left-moving fermions, resulting in a $\cos \left(2 \phi \right)$ term in bosonized form. On the other hand, backscattering of fermions with opposite spin is described by the four-fermion operator 
\begin{equation*}
\psi_{L \uparrow}^{\dagger} \psi_{R \uparrow}^{\phantom{\dagger}} \psi_{R \downarrow}^{\dagger} \psi_{L \downarrow}^{\phantom{\dagger}} + \text{H.c.},
\end{equation*} 
which gives the $\cos \left(2 \sqrt{2} \phi_{\sigma} \right)$ term considered here. This cosine term has a different scaling dimension than $\cos(2 \phi)$, resulting in a different RG equation for the coupling $y_{1 \perp}$ and therefore a different critical $K_{\sigma}$ for the transition from relevant to irrelevant backscattering. Namely, in the case of spinless fermions, the critical value is $K = 1$, meaning that for noninteracting leads the impurity potential is marginal in renormalization and does not change with increasing length scale. For a nonzero potential strength, one would therefore not arrive at perfect conductance at $T = 0$.

\subsection{Strong coupling $L_{\Delta}(l^*) \ll L \ll L_T$: three-step renormalization group}
\label{sec:strong_coupling}

If criterion ii) of Sec.~\ref{sec:renormalization_of_parameters} is reached, there is a finite spin gap while the wire length is finite. Since the RG equations are not valid beyond ${y_{1 \perp}(l^*) \sim 1}$, the renormalization cannot be continued up to a length scale where the cutoff would reach the wire length. One can however assume that also in this situation the behavior of the finite-length wire is the same as that of a local backscattering in a noninteracting wire. In this case, a spin current $\sqrt{2}/\pi \partial_t \phi_{\sigma}(t)$ is generated by the tunneling of instantons: the field $\phi_{\sigma}(t)$ is fixed to a minimum of $\cos\left( 2\sqrt{2} \phi_{\sigma}(t) \right)$ most of the time, but can tunnel from one minimum to the next and create a small but finite current. In such a tunneling event, the argument of the cosine in Eq.~(\ref{eq:point_impurity}) changes by $2 \pi$. Instead of $y_{1 \perp}$, one can find a renormalization equation for a parameter which describes the probability of tunneling -- when $y_{1 \perp} > 1$, the fugacity $f = e^{-S_{\text{inst}}}$ can be used as a perturbative parameter. Here, $S_{\text{inst}}$ is the action of instantons (see Appendix~\ref{app:strong_impurity}) and has the dependence $S_{\text{inst}} \propto \sqrt{y_{1 \perp}}$~\cite{Giamarchi_one_dimension2003}. The RG equation for $f$ is obtained as
\begin{equation}
\frac{d f(l)}{dl} = \left( 1 - \frac{1}{2 K_{\sigma}} \right) f(l) = \frac{1}{2} f(l),
\label{eq:RG_fugacity}
\end{equation}
while $K_{\sigma}$ in the leads is unchanged in renormalization and has the value $K_{\sigma} = 1$. The RG flow corresponding to Eq.~(\ref{eq:RG_fugacity}) is the opposite of $y_{1 \perp}$, as shown in Fig.~\ref{fig:RG_flow_backscattering_fugacity}: For $K_{\sigma} < 1/2$, $f$ flows to zero and for $K_{\sigma} > 1/2$, to one.

The value of $f$ at scale $l = 0$ can in principle be found by matching the value of conductance at $T = T_L$ to the value $\sim e^{-L/L_{\Delta}}$ expected for $T_L < T < \Delta$, as discussed in Sec.~\ref{sec:intermediate_temperature}. We illustrate this procedure numerically in Sec.~\ref{sec:prediction}. According to Eq.~(\ref{eq:RG_fugacity}), $f(l)$ will then grow with increasing length scale. Here, one should note that $l$ in the RG equations is a length scale in both space and imaginary time, $(x, v_{\sigma}\tau)$, and in the renormalization of $f(l)$, the initial cutoff on $v_{\sigma} \tau$ is $\alpha(l = 0) = L$. Similar to Sec.~\ref{sec:renormalization_of_parameters}, one iterates Eq.~(\ref{eq:RG_fugacity}) up to the length scale at which either i) $\alpha(l^*) \sim L_T$ or ii) $f(l^*) \sim 1$. If criterion i) is reached while $f(l^*) < 1$, one can calculate the conductance perturbatively in $f$. A second-order perturbation expansion gives the dependence
\begin{equation}
G_{\sigma}(T) \propto f^2(l^*) \left(\frac{T}{\Lambda} \right)^{4 K_{\sigma} - 2} = f^2(l^*) \left(\frac{T}{\Lambda} \right)^2.
\label{eq:fugacity_scaling}
\end{equation}

The lower the temperature is, the larger is $L_T$ and the further $f$ and $\alpha$ are renormalized. At very low temperature, therefore, $f(l)$ will reach the value 1 before ${\alpha(l) \sim L_T}$, at which point Eq.~(\ref{eq:RG_fugacity}) is not valid anymore. The coupling $y_{1 \perp}(l)$ on the other hand has been renormalized to $y_{1 \perp} < 1$ and can again be used as a perturbative parameter. One therefore switches from Eq.~(\ref{eq:RG_fugacity}) to Eq.~(\ref{eq:RG_point_impurity}) and continues the renormalization of $\alpha$ up to $L_T$. When $y_{1 \perp}$ is perturbative, we have again the result $G_{\sigma} - 1 \propto - y_{1 \perp}^2(l^*) \left(T/\Lambda \right)^2$ as in Sec.~\ref{sec:weak_coupling}. One sees now that at $T \rightarrow 0$, the spin conductance approaches 1. The nonmonotonic dependence of $G_{\sigma}$ on temperature is illustrated in Fig.~\ref{fig:sketch}.
A similar behavior was predicted for charge conductance in the case of umklapp scattering~\cite{Ponomarenko_Mott_insulator1998}.

\section{Temperature dependence of the spin conductance for experimental parameters}
\label{sec:prediction}

\begin{figure}[h]
\centering
\includegraphics[width=\linewidth]{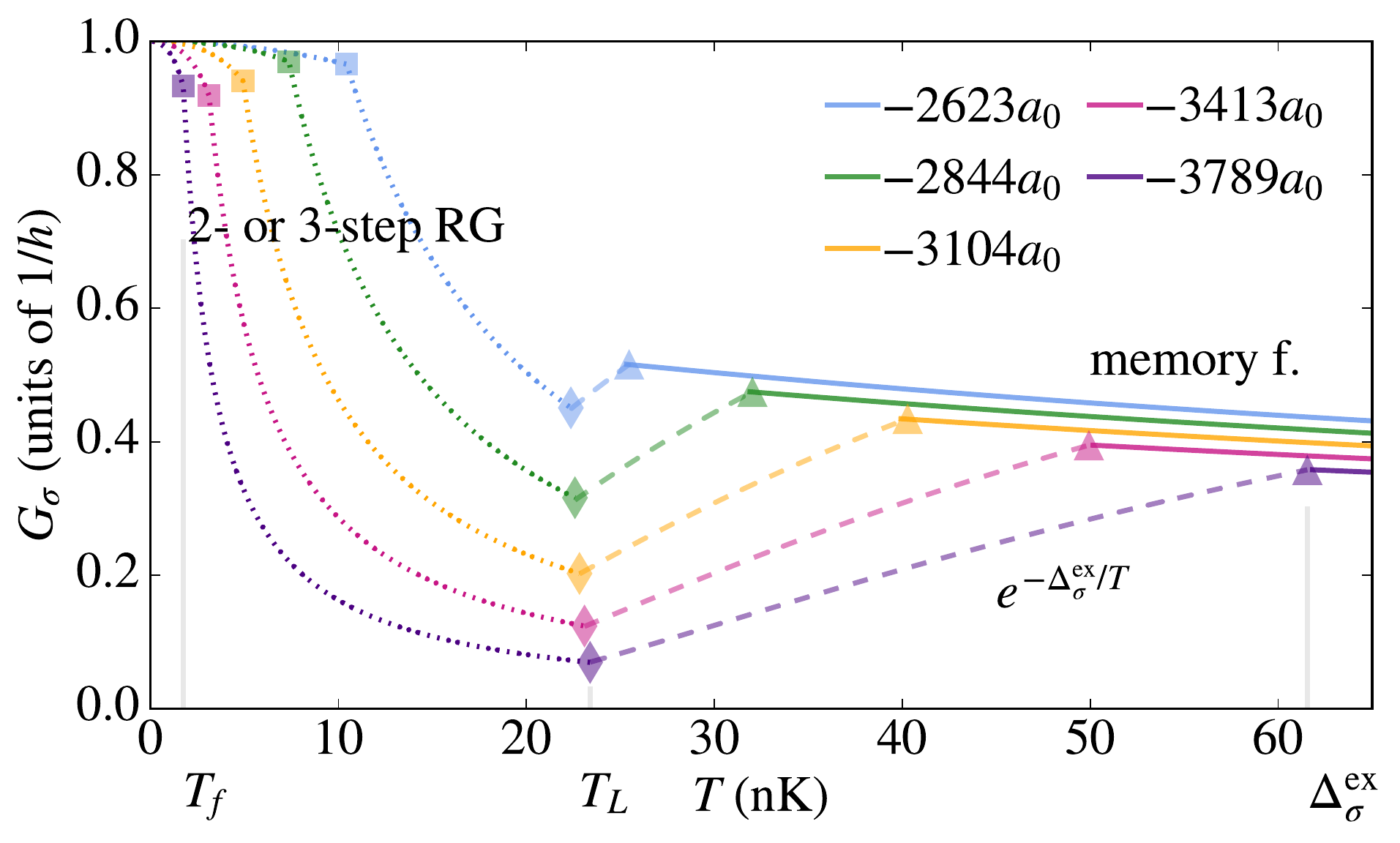}
\caption{The spin conductance as a function of temperature as determined by the various methods described in Secs.~\ref{sec:high_temperature}--\ref{sec:low_temperature}. The different methods are denoted by different line styles: the solid lines corresponds to the memory-function calculation, the dashed lines to the exponential dependence in the intermediate-temperature regime, and the dotted lines to the 2- or 3-step RG results. The colors correspond to the same values of the scattering length $a$ as in Figs.~\ref{fig:memory_function} and~\ref{fig:experimental}. The particle density and is fixed to $\rho_0 = 0.8/\mu$m. The other parameters are given in Appendix~\ref{app:experimental_parameters}. The spin gap given by Eq.~(8) in Ref.~\onlinecite{Fuchs_exactly_solvable2004} is marked with triangles and denoted by~$\Delta_{\sigma}^{\text{ex}}$, which is indicated for the largest negative scattering length. The temperature $T_L$ is marked with diamonds. The temperature at which the fugacity $f$ becomes nonperturbative (see text) is marked with squares and indicated by $T_f$ for the largest negative scattering length. We connect the different expressions for $G_{\sigma}(T)$ in each region by matching the value of the spin conductance at the transition temperatures $\Delta_{\sigma}^{\text{ex}}$, $T_L$, and $T_f$, even though at these values the methods are not strictly valid.}
\label{fig:G_temperature_dependence_experimental}
\end{figure}

The analyses of the previous sections can be used to construct the temperature dependence of the spin conductance for experimentally relevant parameters. We use here the same parameters as in Figs.~\ref{fig:memory_function} and \ref{fig:experimental} and Appendix~\ref{app:experimental_parameters}, apart from the varying temperature. The particle density $\rho_0 = 0.8/\mu$m is fixed to a value which corresponds to the strong-coupling regime $L_{\Delta} < L$ for all scattering lengths used here. Figure~\ref{fig:G_temperature_dependence_experimental} shows the temperature dependence of $G_{\sigma}$ for these parameters, and can be compared to the schematic drawing of Fig.~\ref{fig:sketch}(b). 

At $T > \Delta_{\sigma}$, the spin conductance is calculated by using the memory-function expression for spin conductivity as in Sec.~\ref{sec:memory_function_comparison}. The solid lines on the right side of Fig.~\ref{fig:G_temperature_dependence_experimental} show the result of the memory-function calculation for parameters which are renormalized up to~$\alpha(l^*) = L_T$. The solid lines extend down to the temperature corresponding to the exact value of the spin gap. The spin gap has an exact expression for the Gaudin-Yang model, given by Eq.~(8) in Ref.~\cite{Fuchs_exactly_solvable2004}. This model describes the experiment of Ref.~\onlinecite{Krinner_spin_and_particle2016} in the case of low particle densities where the spin resistance is dominated by the wire and superfluidity in the leads is not a significant factor. 
The spin gap given by Eq.~(8) in Ref.~\cite{Fuchs_exactly_solvable2004} is denoted by $\Delta_{\sigma}^{\text{ex}}$ and marked with triangles in Fig.~\ref{fig:G_temperature_dependence_experimental}. As discussed in Sec.~\ref{sec:memory_function}, the memory-function calculation gives the dependence $G_{\sigma}(T) - 1 \propto -y_{1 \perp}^2 (L/\alpha) \left( T/\Lambda \right)^{4 K_{\sigma} - 3}$. Note that while the schematic diagram of Fig.~\ref{fig:sketch}(b) shows the case $K_{\sigma} < 3/4$ where $G_{\sigma}$ decreases with decreasing temperature, the parameters and the temperature range shown here correspond to $K_{\sigma}(l = 0), K_{\sigma}(l^*) > 3/4$, which leads to an initial increase of $G_{\sigma}$ for decreasing temperature. Continuing the lines down to sufficiently low temperatures would lead to $K_{\sigma}(l^*) < 3/4$ and a downturn of the conductance curve. 
The initial ($l = 0$) and renormalized ($l = l^*$) parameters $y_{1 \perp}$ and $K_{\sigma}$ are shown in Table~\ref{tab:initial_values} for the particle density $\rho_0 = 0.8/\mu$m and temperature $T = 65$ nK, which is the highest temperature included in Fig.~\ref{fig:G_temperature_dependence_experimental}. The parameter values at $l = 0$ are obtained from the experimental parameters as explained in Appendix~\ref{app:experimental_parameters}, and renormalized according to Eq.~(\ref{eq:RG_equations}) as discussed in Sec.~\ref{sec:high_temperature}. Here, the thermal length $L_T$ is the limiting length in the renormalization.

\begin{table}[h]
  \begin{center}
    \caption{Parameters $y_{1 \perp}(l = 0)$ and $K_{\sigma}(l = 0)$ corresponding to the scattering lengths $a$ and particle density $\rho_0 = 0.8/\mu$m used in Fig.~\ref{fig:G_temperature_dependence_experimental}, calculated from Eqs.~(\ref{eq:g_a})--(\ref{eq:separatrix_K}). The renormalized parameters $y_{1 \perp}(l^*)$ and $K_{\sigma}(l^*)$, obtained as explained in Sec.~\ref{sec:renormalization_of_parameters}, are also shown for temperature $T = 65$~nK.}
    \begin{tabular}{c c c c c}
    $a/a_0$			&$y_{1 \perp}(l = 0)$	&$y_{1 \perp}(l^*)$ &$K_{\sigma}(l = 0)$	&$K_{\sigma}(l^*)$  \\
    \hline
      $-2623$		&0.35	&0.41	&0.84	&0.82   \\
      $-2844$		&0.37	&0.43	&0.83	&0.81   \\
      $-3104$ 		&0.39	&0.46	&0.82	&0.80   \\
      $-3413$ 		&0.41	&0.49	&0.81	&0.78   \\
      $-3789$ 		&0.43	&0.52	&0.80	&0.77   \\
      \hline
  \label{tab:initial_values}
    \end{tabular}
  \end{center}
\end{table}

For $T_L < T < \Delta_{\sigma}$, the spin conductance has an exponential form. We use the exact value of the spin gap $C e^{-\Delta_{\sigma}^{\text{ex}}/T}$ where the constant $C$ is fixed by ${G_{\sigma}(T = \Delta_{\sigma}^{\text{ex}})}$ given by the memory-function expression, so that $G_{\sigma}(T) = G_{\sigma}(\Delta_{\sigma}^{\text{ex}}) e^{1 - \Delta_{\sigma}^{\text{ex}}/T}$. For $T < T_L$, there is again a different behavior of the spin conductance when the thermal length exceeds the length of the wire, and we use Eqs.~(\ref{eq:fugacity_scaling}) and~(\ref{eq:perturbative_y1}) to calculate the conductance, as explained in Sec.~\ref{sec:strong_coupling}. At $T = T_L$, the system can be thought of as zero dimensional with $\alpha(l = 0) = L$, so that $\Lambda = v_{\sigma}/L$. We renormalize the fugacity according to Eq.~(\ref{eq:RG_fugacity}), with the value $f(l = 0)$ fixed by $f(l = 0) = \sqrt{G_{\sigma}(T_L)}$. For temperatures at which the cutoff reaches the thermal length $\alpha(l^*) = L_T$ before $f(l) = 1$, we compute the conductance by Eq.~(\ref{eq:fugacity_scaling}). For sufficiently large $L_T$, $f(l^*) = 1$ is reached first. In this case, we switch to the RG equation~(\ref{eq:RG_point_impurity}) for $y_{1 \perp}$ and calculate the conductance from Eq.~(\ref{eq:perturbative_y1}) using $y_{1 \perp}(l = 0) = (\Lambda/T)\sqrt{1 - G_{\sigma}(T_f)}$, where $T_f$ denotes the temperature at which $f(l^*)$ becomes nonperturbative and $\Lambda = v_{\sigma}/\alpha(l^*)$ is the value at the corresponding length scale. Note that while we have connected the perturbative and exponential expressions for $G_{\sigma}(T)$ to make it a continuous function of temperature, these results are not strictly valid at $T = \Delta_{\sigma}$, $T_L$, or $T_f$.

\section{Discussion}
\label{sec:discussion}

The RG arguments together with perturbative and nonperturbative calculations show that a finite-length wire connected to noninteracting leads has conducting and insulating phases depending on the relative magnitudes of the length scales $L$, $L_{\Delta}$, and $L_T$, or conversely energy scales $T_L$, $\Delta_{\sigma}$, and $T$. At high temperature $T \gg T_L, \Delta_{\sigma}$, the spin conductance has the correction ${G_{\sigma} - 1 \propto -y_{1 \perp}^2 (L/\alpha) (T/\Lambda)^{4 K_{\sigma} - 3}}$ given by the memory-function calculation (see Appendix~\ref{app:memory_function}). We find that the memory-function result agrees well with experimental data in the region where it is expected to be valid.

In the case of weak coupling, $\Delta_{\sigma} < T_L$, we find the dependence ${G_{\sigma} - 1 \propto -y_{1 \perp}^2 (T/\Lambda)^2}$ in the low-temperature regime where $T < T_L$. 
In the case of strong coupling $T_L < \Delta_{\sigma}$, the spin conductance behaves as $e^{-\Delta_{\sigma}/T}$ when temperature falls below the spin gap. A renormalization-group analysis shows that the conductance starts to grow again for $T < T_L$ and the system approaches perfect conductance at $T \rightarrow 0$. A similar temperature dependence was predicted for the charge conductance in the case of umklapp scattering in a Mott-Hubbard insulator \cite{Ponomarenko_treshold_features1997, Ponomarenko_Mott_insulator1998}. It is interesting to note that the result differs from the point-like barrier considered by Kane and Fisher~\cite{Kane_transport1992, Kane_transmission1992}: the backscattering of spinless fermions from an impurity potential is described by a two-fermion operator whereas the backscattering of fermions with opposite spins considered here is described by a four-fermion operator. This leads to a different set of RG equations and a vanishing coupling for the backscattering of fermions with opposite spin at $T = 0$. In the case of spinless fermions, in contrast, the system does not reach perfect conductance at $T = 0$ since the impurity potential is marginal in renormalization and does not vanish. 

Unlike the static spin conductivity which is zero in the gapped phase, the spin conductance of a finite-length wire is perfect at $T = 0$ for either repulsive or attractive interactions. This reflects the fundamental difference of these quantities. The static conductivity, or Drude weight, can be measured from the persistent current on a ring threaded by a flux, whereas conductance measures the current response of a linear system attached to reservoirs.
We consider a system with spin-charge separation, where the charge degree of freedom is described by a quadratic model and has a conductance equal to the conductance quantum. The spin drag in this case is $\Gamma = (1 - G_{\sigma})/2$. We therefore find that at zero temperature, there is no spin drag for either attractive or repulsive interactions. However, a considerable spin drag exists in the intermediate temperature regime.

The nonmonotonic temperature dependence of spin conductance, or conversely spin drag, could be measured in a setup similar to Refs.~\onlinecite{Krinner_spin_and_particle2016, Lebrat_band_and_correlated2018, Corman_dissipative_point_contact2019}, and we have illustrated the behavior of the conductance for parameters similar to those in Ref.~\onlinecite{Krinner_spin_and_particle2016}. The temperature in the experiment of Ref.~\onlinecite{Krinner_spin_and_particle2016} is determined to be 55 nK, whereas the increase of spin conductance at low temperature occurs only below $T_L \approx 22$ nK for the parameters used in Fig.~\ref{fig:G_temperature_dependence_experimental}. As the temperature $T_L = v_{\sigma}/L$ increases with decreasing wire length, a shorter wire could allow for observing the upturn of the conductance at $T < T_L$. For a very short system and a low particle density, however, a field-theory description may not be accurate. Lower temperatures in the experiment could therefore help to observe the nonmonotonic temperature dependence unambiguously.

\section{Conclusions}
\label{sec:conclusions}

In summary, we have investigated theoretically the spin conductance in an interacting one-dimensional wire connected to noninteracting leads. We use the sine-Gordon model as a low-energy field-theory description of the spin degree of freedom in the wire. Spin transport is an excellent way to study the sine-Gordon model since a cosine term arises naturally from the backscattering of fermions with opposite spin. Therefore, there is no need to introduce an external periodic potential~\cite{Lebrat_band_and_correlated2018}. The temperature dependence of the spin conductance is found in different parameter regimes by combining renormalization-group arguments with perturbative results and instanton calculations. The spin conductance is found to have a nonmonotonic dependence on temperature in a region of parameters which mostly coincides with the strong-coupling regime, as sketched in Fig.~\ref{fig:sketch}(b). We provide an estimate for the temperature range where the nonmonotonic dependence could be observed experimentally. We compare our theoretical results to experimental data from measurements with cold atoms in a wire connected to reservoirs~\cite{Krinner_spin_and_particle2016}, and find a good qualitative agreement.

The present analysis suggests several developments. On the experimental front, as discussed in the previous section, observations are at the moment limited to spin-rotation-invariant interactions and the high-temperature regime where the thermal length is the shortest length scale. It would be interesting to access also other regimes. For the spin transport, this could be achieved by spin-dependent interactions. Alternatively, one could study charge transport in a lattice potential at half-filling, which would allow one to tune independently the parameters corresponding to Fig.~\ref{fig:RG_flow}. Such systems are however more sensitive to the inhomogeneity created by the confining potential. Spin transport with spin-dependent interactions therefore seems like a promising route.

On the theory side, it would be interesting to study the present problem by microscopic methods, for example a numerical analysis by methods such as DMRG or other tensor-network algorithms~\cite{Brenes_tensor_network2019}. This could help in obtaining more quantitative predictions. An important theoretical question is the precise effect of temperature. In particular, we have considered temperature as a cutoff, and it is not clear what happens to spin transport if this assumption is not valid. Numerical analyses or studies based on exactly solvable models~\cite{Mehta_nonequilibrium_transport2006} would be useful in answering this question. Last but not least, we consider here the spin conductance in the linear-response regime. It is well known that for systems such as an impurity in a Tomonaga-Luttinger liquid, the nonlinear-response regime is nontrivial and reflects the power-law correlations in the TLL. An interesting prospect is therefore to study the consequences of the various energy scales considered here on the nonlinear response.

\begin{acknowledgments}
We thank Tilman Esslinger, Philipp Fabritius, Jeffrey Mohan, Michele Filippone, and Pjotrs Grisins for useful and inspiring discussions, and Christophe Berthod and Jo\~ao Ferreira for feedback on the manuscript. This work was supported by the Swiss National Science Foundation under Division II and by the ARO-MURI Non-equilibrium Many-body Dynamics grant (W911NF-14-1-0003). We also acknowledge the Swiss National Science Foundation (Grants No. 182650 and No. NCCR-QSIT) and European Research Council advanced grant TransQ (GrantNo. 742579) for funding. L.C. is supported by an ETH Zurich Postdoctoral Fellowship, the Marie Curie Actions for People COFUND program, and the European Union Horizon 2020 Marie Curie TopSpiD program (Grant No. 746150). The calculations were performed in the University of Geneva with the clusters Mafalda and Baobab.
\end{acknowledgments}

\appendix

\section{Cross-conductivity}
\label{app:conductivity}

The linear response relation defining the conductivity can be written as $I = \sigma E$, where $I$ denotes current, $\sigma$ conductivity, and $E$ the field. In the case of a spin-dependent field, the response is given by the conductivity matrix
\begin{equation}
\begin{pmatrix}
I_{\uparrow} \\
I_{\downarrow}
\end{pmatrix}
=
\begin{pmatrix}
\sigma_{\uparrow \uparrow} 		&\sigma_{\uparrow \downarrow}  \\
\sigma_{\downarrow \uparrow} 	&\sigma_{\downarrow \downarrow}
\end{pmatrix}
\begin{pmatrix}
E_{\uparrow} 	\\
E_{\downarrow}
\end{pmatrix}.
\end{equation}
The cross-conductivity $\sigma_{\uparrow \downarrow} = \sigma_{\downarrow \uparrow}$ gives the response of spin-up current to a field that is applied on spin-down particles. This corresponds to spin drag in the $L \rightarrow \infty$ limit. It can be written as the current-current correlation
\begin{equation*}
\sigma_{\uparrow \downarrow}(k, \omega) = -\frac{i}{\omega}\braket{j_{\uparrow}^*(k, \omega) j_{\downarrow}(k, \omega)},
\end{equation*}
which can be evaluated in terms of correlations of bosonic fields.

Current is related to the bosonic fields as 
\begin{equation*}
j_{\uparrow, \downarrow}(x, t) = \frac{1}{\pi} \partial_t \phi_{\uparrow, \downarrow}(x, t).
\end{equation*}
which in terms of the spin and charge fields can be written as
\begin{align*}
j_{\uparrow}(x, t) &= \frac{1}{2} \left[ j_{\rho}(x, t) + j_{\sigma}(x, t) \right] \\
&= \frac{1}{\sqrt{2}\pi} \left[ \partial_t \phi_{\rho}(x, t) + \partial_t \phi_{\sigma}(x, t) \right] \\
j_{\downarrow}(x, t) &= \frac{1}{\sqrt{2}\pi} \left[ \partial_t \phi_{\rho}(x, t) - \partial_t \phi_{\sigma}(x, t) \right].
\end{align*}
The cross-conductivity is given by the Kubo formula \cite{Mahan_many-particle2000}
\begin{align}
\begin{split}
\sigma_{\uparrow \downarrow}(x, t) &= \frac{1}{i(\omega + i \delta)} \cdot \frac{1}{2 \pi^2} \\
\times &\int_{L/2}^{L/2} dx' dt' \Big[ \braket{\partial_t \phi_{\rho}(x, t); \partial_{t'} \phi_{\rho}(x', t')} \\
&- \braket{\partial_t \phi_{\sigma}(x, t); \partial_{t'} \phi_{\sigma}(x', t')} \Big] \\
&= \sigma_{\rho}(x, t) - \sigma_{\sigma}(x, t),
\end{split}
\label{eq:current-current_correlation}
\end{align}
where 
\begin{equation*}
\braket{A(x, t);B(x', t)} = -i \Theta(t - t')\braket{ \{ A(x, t), B(x', t') \} }
\end{equation*}
denotes the retarded correlation function and $\Theta(t)$ the step function.
In the case of free particles, the two terms in Eq.~(\ref{eq:current-current_correlation}) cancel and there is no spin drag, whereas for interacting particles, the terms do not necessarily cancel.

\subsection{Attractive interactions}

For attractive interactions, the cosine term of Eq.~(\ref{eq:sine-gordon}) is relevant. One can approximate the cosine by a Taylor expansion close to the minimum, which for $g_1 < 0$ is at $\phi_{\sigma} = 0$. Up to the first two terms,
\begin{equation*}
\cos \left( \sqrt{8} \phi_{\sigma} \right) \approx 1 - \frac{1}{2} \left( \sqrt{8} \phi_{\sigma} \right)^2 = 1 - 4 \phi_{\sigma}^2.
\end{equation*}
Leaving out the constant term, the spin Hamiltonian becomes
\begin{equation*}
H_{\sigma} = \frac{1}{2 \pi} \left[ v_{\sigma} K_{\sigma} \left( \partial_x \theta_{\sigma} \right)^2 + \frac{v_{\sigma}}{K_{\sigma}} \left( \partial_x \phi_{\sigma} \right)^2 \right] - \frac{2 g_1}{(\pi \alpha)^2} \phi_{\sigma}^2.
\end{equation*}
The expectation values in Eq.~(\ref{eq:current-current_correlation}) can be evaluated as functional integrals. For the charge conductivity, one obtains
\begin{align*}
\sigma_{\rho}(\omega) &= -\frac{i}{2 \pi^2}(\omega + i\delta) \braket{\phi_{\rho}^*(k = 0, \omega_n) \phi_{\rho}(k = 0, \omega_n)}_{i \omega_n \rightarrow \omega + i\delta} \\
&= \frac{i}{2 \pi} \frac{v_{\rho} K_{\rho}}{\omega + i \delta} \\
&= \frac{1}{2} v_{\rho} K_{\rho} \delta(\omega) + \frac{i}{2 \pi} v_{\rho} K_{\rho} P \left( \frac{1}{\omega} \right),
\end{align*}
where $P$ denotes the principal value. At $\omega = 0$, the charge conductivity is therefore infinite. For the spin conductivity,
\begin{equation}
\sigma_{\sigma}(\omega) = \frac{i}{2 \pi} \cdot \frac{\omega v_{\sigma} K_{\sigma}}{\omega^2 - \frac{4 g_{1 \perp} v_{\sigma} K_{\sigma}}{\pi \alpha^2} + i\delta},
\end{equation}
so that $\sigma_{\sigma}(\omega = 0) = 0$. The cross-conductivity $\sigma_{\uparrow \downarrow}(\omega)$ is therefore given by only the charge conductivity at $\omega \rightarrow 0$ (static electric field), and is nonzero. Physically, a nonzero spin drag for attractive interactions is caused by the spin-up and spin-down particles forming pairs.

\subsection{Repulsive interactions}

For $K_{\sigma} > 1$, below the separatrix in Fig.~\ref{fig:RG_flow}, the coupling $y_{1 \perp}$ renormalizes to zero and the cosine term of Eq.~(\ref{eq:sine-gordon}) is irrelevant. The Luttinger parameter $K_{\sigma}$ renormalizes to a value $K_{\sigma}^*$ on the $y_{1 \perp} = 0$ axis. For repulsive spin-rotation invariant interactions on the separatrix, $K_{\sigma}^* = 1$. The charge conductivity is the same as in the case of attractive interactions, and the spin conductivity for the renormalized quadratic model has the same form as the charge conductivity. We therefore have
\begin{align*}
\sigma_{\uparrow \downarrow}(\omega) &= \frac{i}{2 \pi} \left( \frac{v_{\rho} K_{\rho}}{\omega + i\delta} - \frac{v_{\sigma} K_{\sigma}^*}{\omega + i \delta} \right) \\
&= \frac{1}{2} (v_{\rho} K_{\rho} - v_{\sigma} K_{\sigma}^*) \left[\delta(\omega) + \frac{i}{\pi} P \left( \frac{1}{\omega} \right) \right].
\end{align*}
For a Galilean-invariant system, $v_{\rho} K_{\rho} = v_F$. On the separatrix, one has the renormalized value $K_{\sigma}^* = 1$, and the expression simplifies to
\begin{equation*}
\sigma_{\uparrow \downarrow}(\omega) = \frac{1}{2} (v_F - v_{\sigma}) \left[\delta(\omega) + \frac{i}{\pi} P \left( \frac{1}{\omega} \right) \right].
\end{equation*}
At $\omega = 0$, the cross-conductivity is given by the delta-function term. If $v_F \neq v_{\sigma}$, this expression is nonzero, giving a finite spin drag.

\section{Conductivity at high temperature}
\label{app:memory_function}

Conductivity at high temperature with respect to the gap can be calculated by writing the conductivity in terms of the memory function $M(\omega, T)$ \cite{Giamarchi_umklapp_resistivity1991}. This function contains the spin current-current correlation, which can be calculated perturbatively. For fermions with spin, the Kubo formula for conductivity can be written in the form 
\begin{equation*}
\sigma_{\sigma}(\omega) = \frac{i}{\omega} \left[ \frac{2 v_{\sigma} K_{\sigma}}{\pi} + \chi(\omega) \right],
\end{equation*}
which gives the spin conductivity in terms of the retarded spin current--current correlation $\chi(\omega)$. At $\omega = 0$, one obtains $\chi(0) = -2 v_{\sigma} K_{\sigma}/\pi$, so that the spin conductivity can be written as
\begin{equation*}
\sigma_{\sigma}(\omega) = \frac{i 2 v_{\sigma} K_{\sigma}}{\pi \omega} \left[ 1 - \frac{\chi(\omega)}{\chi(0)} \right].
\end{equation*}
In terms of the memory function $M(\omega, T)$,
\begin{equation}
\sigma_{\sigma}(\omega, T) = \frac{i 2 v_{\sigma} K_{\sigma}}{\pi} \cdot \frac{1}{\omega + M(\omega, T)},
\label{eq:conductivity_memory_function}
\end{equation}
where
\begin{equation*}
M(\omega) = \frac{\omega \chi(\omega)}{\chi(0) - \chi(\omega)}.
\end{equation*}
When $y_{1 \perp}$ is small, $\chi(\omega \neq 0)$ is small and one can approximate the denominator as $\chi(0) - \chi(\omega) \approx \chi(0)$. The numerator can be expressed as \cite{Giamarchi_umklapp_resistivity1991}
\begin{equation*}
\omega \chi(\omega) = -\frac{1}{\omega}\left[ \braket{F; F}_{\omega} - \braket{F; F}_{\omega = 0} \right]
\end{equation*}
where $F$ is the commutator $F = [H, j(t)]$ and
\begin{align}
\braket{F; F}_{\omega} = -i \int_0^{\infty} dt e^{i \omega t} \braket{[F(t), F(0)]}
\label{eq:F_expression}
\end{align}
For $\omega = 0$, one has $\braket{F; F}_{\omega = 0} = -i \int_0^{\infty} dt \braket{[F(t), F(0)]}$. 

The expectation value cannot be evaluated for the full Hamiltonian with the backscattering term, but for small $y_{1 \perp}$we can approximate it by using the quadratic Hamiltonian $H_{\sigma}^0$. The memory function is thus approximated as
\begin{equation*}
M(\omega) \approx -\frac{\braket{F; F}_{\omega}^0 - \braket{F; F}_{\omega = 0}^0}{\omega \chi(0)}.
\end{equation*}
At $T \gg \Delta_{\sigma}$ and $\omega \ll T$, one obtains the expression~\cite{Giamarchi_umklapp_resistivity1991}
\begin{align}
\begin{split}
M(\omega = 0, T) \approx i \frac{g_{1 \perp}^2 K_{\sigma}}{\pi^3 \alpha^2} &B^2 (K_{\sigma}, 1 - 2 K_{\sigma}) \cos^2 (\pi K_{\sigma}) \\
&\times \frac{1}{T} \left( \frac{2 \pi \alpha T}{v_{\sigma}} \right)^{4 K_{\sigma} - 2},
\end{split}
\label{eq:memory_function}
\end{align}
where $B$ denotes the beta function. At $\omega = 0$, Eq.~(\ref{eq:conductivity_memory_function}) gives the expression
\begin{equation}
\sigma_{\sigma}(0) = \frac{i 2 v_{\sigma} K_{\sigma}}{\pi M(0)}.
\label{eq:conductivity_at_omega0}
\end{equation}
for the spin conductivity. Substituting the expression (\ref{eq:memory_function}) shows that the temperature dependence has the form
\begin{equation}
\sigma_{\sigma}(0) \propto \frac{1}{g_{1 \perp}^2} T^{3 - 4 K_{\sigma}}.
\label{eq:temperature_dependence}
\end{equation}

If $K_{\sigma}$ does not change in renormalization, one obtains the same temperature dependence of the spin conductivity with the renormalized coupling
\begin{equation*}
g_{1 \perp}(l^*) = g_{1 \perp}(0) e^{(2 - 2 K_{\sigma}) l^*}
\end{equation*}
where the length scale $l^*$ is given by $\alpha(l^*) = \alpha(0) e^{l^*} = v_{\sigma}/T$. If $K_{\sigma}$ is not scale invariant, one obtains a modified exponent in Eq.~(\ref{eq:temperature_dependence}). The expression for the memory function is valid when the cosine term is irrelevant and a perturbation expansion in $y_{1 \perp}$ converges. One can however use the memory function approach also in the regime where the cosine term is relevant when $\alpha(l)$ reaches $L_T$ before $y_{1 \perp}(l) \sim 1$. This will only occur at energy scales (temperature or frequency) larger than the spin gap.

\section{Experimental parameters}
\label{app:experimental_parameters}

Experimental parameters such as optical trapping frequencies and the scattering length can be related to the effective parameters $K_{\sigma}$ and $y_{1 \perp}$ of Eq.~(\ref{eq:bosonization_hamiltonian}) via the Gaudin-Yang model of Eq.~(\ref{eq:gaudin-yang}). In the Gaudin-Yang model, interactions are described by an effective delta function potential with strength $g_{1 \perp}$. For fermions confined in a quasi-1D geometry, $g_{1 \perp}$ can be calculated as \cite{Fuchs_exactly_solvable2004}
\begin{equation}
g_{1 \perp} = \frac{2 \hbar \omega_{\perp} a}{1 - A \frac{a}{a_{\perp}}},
\label{eq:g_a}
\end{equation}
where $\hbar = h/(2 \pi)$ is the reduced Planck constant, $\omega_{\perp}$ the transversal confinement frequency, $a$ the s-wave scattering length in three dimensions, $a_{\perp}$ the oscillator length, $m$ the mass of the atoms, $A$ a constant coming from a wave-function expansion in the derivation of the effective 1D interaction potential~\cite{Olshanii_atomic_scattering1998}. The transversal confinement frequency is given by $\omega_{\perp} = \sqrt{\omega_x \omega_z}$ and the oscillator length by $a_{\perp} = \sqrt{\hbar/(m \omega_{\perp})}$. The parameters corresponding to Fig.~\ref{fig:experimental} are the same as those of Ref.~\onlinecite{Krinner_spin_and_particle2016} and are listed in Table~\ref{tab:experimental_parameters}.

\begin{table}[h]
  \begin{center}
    \caption{Experimental parameters used in Figs.~\ref{fig:memory_function}, \ref{fig:experimental}, and \ref{fig:G_temperature_dependence_experimental}.}
    \label{tab:experimental_parameters}
    \begin{tabular}{l c r l} 
    Parameter								&Symbol			&Value \\
    \hline
      reduced Planck constant				&$\hbar$		&1.054$\times 10^{-34}$ &Js \\
      Boltzmann constant					&$k_B$			&1.38$\times 10^{-23}$ &J/K \\
      Bohr radius 							&$a_0$			&5.25$\times 10^{-11}$ &m  \\
      confinement freq. in $x$ direction &$\omega_x/(2 \pi)$ 	&23.2 &kHz  \\
      confinement freq. in $z$ direction &$\omega_z/(2 \pi)$ 	&9.2 &kHz  \\
      transversal confinement frequency		&$\omega_T/(2 \pi)$		&14.4  &kHz\\
      ${}^6$Li mass							&$m$			&6 &amu \\
      atomic mass unit						&amu			&1.66$\times 10^{-27}$ &kg \\
      constant								&$A$			&1.0326  \\
      oscillator length						&$a_{\perp}$	&1.1$\times 10^{-5}$ &m\\
      wire length							&$L$			&5.5$\times 10^{-6}$ &m \\
      temperature							&$T$			&5.5$\times 10^{-8}$ &K \\ 
      \hline
    \end{tabular}
  \end{center}
\end{table}

The dimensionless coupling $y_{1 \perp}$ is given by 
\begin{equation}
y_{1 \perp} = \frac{|g_{1 \perp}|}{\pi v_{\sigma}},
\label{eq:y_g}
\end{equation}
where we take the absolute value of $g_{1 \perp}$ to limit the analysis to $y_{1 \perp} > 0$ as discussed in Sec.~\ref{sec:renormalization_group_gapped_and_gapless}. For the Gaudin-Yang model, the spin velocity $v_{\sigma}$ has the analytic expressions
\begin{align}
\frac{v_{\sigma}}{v_F} \simeq
\begin{cases}
1 - \frac{\gamma}{\pi^2} + \dots, &\frac{1}{\gamma} \rightarrow -\infty \\
-\frac{\gamma}{\pi \sqrt{2}} \left( 1 - \frac{2}{\gamma} + \cdots \right), &\frac{1}{\gamma} \rightarrow 0^-
\end{cases}
\label{eq:spin_velocity}
\end{align}
in the limits of strong and weak attractive interactions~\cite{Fuchs_exactly_solvable2004}. Here, $\gamma$ is the dimensionless parameter $\gamma = m g_{1 \perp}/(\hbar^2 \rho_0)$. In practice, we interpolate between the weak- and strong-interaction expressions to compute both the spin velocity of Eq.~(\ref{eq:spin_velocity}) and the exact spin gap by Eq.~(8) in Ref.~\onlinecite{Fuchs_exactly_solvable2004}. At $K = 1$, the spin velocity is equal to the Fermi velocity $v_F = \hbar \rho_0 \pi/(2 m)$, where $\rho_0$ is the particle density. The Luttinger parameter $K_{\sigma}$ is fixed by $y_{1 \perp}$ on the separatrix:
\begin{equation}
K_{\sigma} = \sqrt{\frac{1 - \frac{y_{1 \perp}}{2}}{1 + \frac{y_{1 \perp}}{2}}}.
\label{eq:separatrix_K}
\end{equation}
We choose the short-distance cutoff to be equal to the inverse density $\alpha = \rho_0^{-1}$. The parameters $y_{1 \perp}(l = 0)$ and $K_{\sigma}(l = 0)$ calculated from Eqs.~(\ref{eq:g_a})--(\ref{eq:separatrix_K}) for the scattering lengths shown in Figs.~\ref{fig:memory_function}, \ref{fig:experimental}, and \ref{fig:G_temperature_dependence_experimental} are given in Table~\ref{tab:initial_values} of the main text.
We use these parameter values as the initial values at $l = 0$ and renormalize them according to Eq.~(\ref{eq:RG_equations}) as discussed in Sec.~\ref{sec:high_temperature}. Figures~\ref{fig:renormalized_parameters_LT} and \ref{fig:renormalized_parameters} show the renormalized values $y_{1 \perp}(l^*)$ and $K_{\sigma}(l^*)$ as functions of the initial ones.

\section{Renormalization of $y_{1 \perp}$ and $K_{\sigma}$ at low temperature $L, L_{\Delta} \ll L_T$}
\label{app:renormalization_of_parameters}

\begin{figure}[h]
\centering
\includegraphics[width=\linewidth]{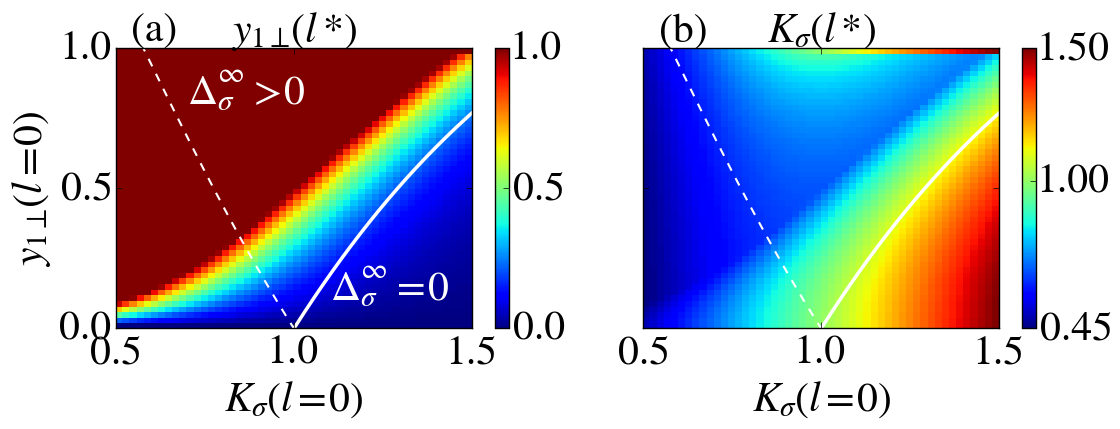}
\caption{The renormalized parameters $y_{1 \perp}(l^*)$ and $K_{\sigma}(l^*)$ as functions of $K_{\sigma}(l = 0)$ and $y_{1 \perp}(l = 0)$. The parameters have been evolved according to Eqs.~(\ref{eq:RG_equations}) up to the length scale $l^*$ at which either $\alpha(l^*) \approx L$ or $y_{1 \perp}(l^*) \approx 1$. The white lines show the separatrix $y_{1 \perp}(0) = 2 \left|K_{\sigma}^2(0) - 1 \right|/ \left|K_{\sigma}^2(0) + 1 \right|$ which corresponds to spin-rotation invariant interactions. The solid white line separates the regions corresponding to a gapped $\Delta_{\sigma} > 0$ and gapless $\Delta_{\sigma} = 0$ system in the limit $L \rightarrow \infty$ as in Fig.~\ref{fig:renormalized_parameters_LT}. We have used here $L = 5.5 \mu$m, $\alpha(l = 0) = \rho_0^{-1} = 1.25 \mu \text{m}$ and the discretization $\delta l = 0.1 \mu$m in Eq.~(\ref{eq:RG_equations}). The same values are used in Fig.~\ref{fig:renormalized_parameters_LT}.}
\label{fig:renormalized_parameters}
\end{figure}

We analyze the spin conductance at low temperature by renormalizing the parameters up to a length scale $l^*$ at which one of two criteria is reached: i) either $\alpha(l^*) \sim L$, at which point the system can be thought of as zero-dimensional and has different RG equations, or ii) $y(l^*) \sim 1$.
Figure~\ref{fig:renormalized_parameters} shows how the renormalized parameters $K_{\sigma}(l^*)$ and $y_{1 \perp}(l^*)$ depend on $K_{\sigma}(l = 0)$ and $y_{1 \perp}(l = 0)$, similar to Fig.~\ref{fig:renormalized_parameters_LT}. 
Note that the diagram for the RG flow in Fig.~\ref{fig:RG_flow} corresponds to the thermodynamic limit, whereas the finite length of the wire in Fig.~\ref{fig:renormalized_parameters} leads to a region with $y_{1 \perp}(l^*) < 1$ above the separatrix.

\section{Renormalization equation for the strong local backscattering}
\label{app:strong_impurity}

When $T < T_L < \Delta_{\sigma}$, renormalization of the parameters leads to $y_{1 \perp}(l^*) \sim 1$ while $\alpha(l^*) < L$. One cannot use RG equations which are perturbative in $y_{1 \perp}$ to renormalize the parameters further. In this situation, even though $\alpha(l^*) < L$, we can expect the conductance of the finite-length wire to be sufficiently well described by a local backscattering term at $x = 0$ with $y_{1 \perp} \gtrsim 1$. This local term in the Hamiltonian is given by Eq.~(\ref{eq:point_impurity}).
Instead of $y_{1 \perp}$, one can now use a different parameter related to local backscattering to formulate perturbative RG equations. 

To find the renormalization equation in the case of a strong local backscattering term, we consider the partition function
\begin{equation*}
Z = \int D \phi_{\sigma}(\tau) e^{-S}.
\end{equation*}
The action $S = S_0 + S_0' + S_g$ contains the quadratic part $S_0$, a term $S_0'$ coming from the regularization of $S_0$ at short times, and the backscattering at $x = 0$
\begin{equation*}
S_g = \frac{y_{1 \perp} \pi v_{\sigma}}{2 \pi^2 L} \int d\tau \cos \left(2 \sqrt{2} \phi_{\sigma}(x = 0, t) \right),
\end{equation*}
When the coefficient of this term is large, one can evaluate the partition function by a saddle-point approximation of the action. The trajectories $\phi_{\sigma}(\tau)$ which minimize the action correspond to instanton solutions of the Euler-Lagrange equation. One can think of the instantons as tunneling events where the field $\phi_{\sigma}$ shifts from one minimum of the cosine to the next. The minima are separated by $2 \pi$, so that $\phi_{\sigma}$ changes by $\pi/\sqrt{2}$ in a tunneling event. 
The instanton-type solutions therefore have the shape of a step or kink in time where $\phi_{\sigma}(\tau)$ changes by $\pi/\sqrt{2}$. 

When there are multiple such steps at times $\tau_i$, the solution can be written as the sum of functions $\tilde{\phi}_{\sigma}(\tau - \tau_i)$ with one step,
\begin{equation*}
\phi_{\sigma}(\tau) = \sum_i \epsilon_i \tilde{\phi}_{\sigma}(\tau - \tau_i),
\end{equation*} 
where $\epsilon_i = \pm 1$ for steps in the positive or negative direction. 
Similar to the spinless case in Refs.~\onlinecite{Kosterlitz_critical_properties1974, Giamarchi_one_dimension2003}, we obtain the partition function for the spin sector as
\begin{align*}
Z = \sum_{p = 0}^{\infty} f^{2 p} &\int_0^{\beta} \frac{d\tau_{2 p}}{\delta} \int_0^{\tau_{2 p - \delta}} \frac{d \tau_{2 p - 1}}{\delta} \dots \int_0^{\tau_2 - \delta} \frac{d\tau_1}{\delta} \\
&\times \sum_{\epsilon_1, \dots \epsilon_{2 p} = \pm} e^{\frac{1}{K_{\sigma}} \sum_{i > j} \epsilon_i \epsilon_j \ln \left( \frac{\tau_i - \tau_j}{\delta} \right)}.
\end{align*} 
Here, the first sum accounts for trajectories with different numbers of instantons $p$. In order to allow only trajectories which are periodic in imaginary time, $\phi_{\sigma}(\beta) = \phi_{\sigma}(0)$, there can only be an even number of instantons on a given trajectory. For the same reason, one has the condition $\sum_i \epsilon_i = 0$. The parameter $f$ is the fugacity of an instanton,
\begin{equation*}
f = e^{-S_{\text{inst}}},
\end{equation*}
containing the instanton action $2 p S_{\text{inst}} = S_0' + S_{g}$. The instanton action can be obtained as a constant ${S_{\text{inst}} \propto \sqrt{y_{1 \perp}}}$. 
When $y_{1\perp} > 1$, $f = e^{-S_{\text{inst}}} \ll 1$, and we can approximate the partition function by the first two terms,
\begin{align*}
Z &= 1 + f^{2} \int_0^{\beta} \frac{d\tau_{2}}{\delta} \int_0^{\tau_2 - \delta} \frac{d\tau_1}{\delta} \sum_{\epsilon_1, \epsilon_{2} = \pm} e^{\frac{1}{K_{\sigma}} \epsilon_1 \epsilon_2 \ln \left( \frac{\tau_2 - \tau_1}{\delta} \right)} \\
&= 1 + 2 f^{2} \int_0^{\beta} \frac{d\tau_{2}}{\delta} \int_0^{\tau_2 - \delta} \frac{d\tau_1}{\delta} \left( \frac{\delta}{\tau_2 - \tau_1} \right)^{\frac{1}{K_{\sigma}}}.
\end{align*}
The integral has the scaling dimension $L^{2 - \frac{1}{K_{\sigma}}}$, which gives the RG equation
\begin{equation*}
\frac{d f(l)}{dl} = \left( 1 - \frac{1}{2 K_{\sigma}} \right) f(l).
\end{equation*}

\bibliographystyle{apsrev4-1-with-titles}
\bibliography{bibfile}

\end{document}